\pdfoutput=1
\documentclass[useAMS]{mn2e}
\usepackage{times}
\usepackage{amssymb}
\usepackage{amsmath}
\usepackage{rotating}
\usepackage{color}
\usepackage{epsfig}
\usepackage{graphicx}
\usepackage{color}
\input{epsf}

\title[UV/optical reprocessing in NGC 5548]
{The origin of the UV/optical lags in NGC 5548}

\author[E. Gardner and C. Done]
{Emma Gardner and Chris Done\\
Centre for Extragalactic Astronomy, Department of Physics, University of Durham, South Road,
Durham DH1 3LE, UK\\}

\date{Submitted to MNRAS}
\pagerange{\pageref{firstpage}--\pageref{lastpage}} \pubyear{}

\def\mnras{MNRAS}
\def\apj{ApJ}
\def\apjl{ApJ}
\def\apjs{ApJS}
\def\aap{A\&A}

\def\nar{New A Rev.}
\def\pasj{PASJ}

\begin{document}

\topmargin = -0.5cm
\maketitle

\label{firstpage}

\begin{abstract}

The new multi-wavelength monitoring campaign on NGC 5548 shows clearly
that the variability of the UV/optical lightcurves lags by
progressively longer times at longer wavelengths, as expected from
reprocessing of an optically thick disk, but that the timescales are
longer than expected for a standard Shakura-Sunyaev accretion disc. We
build a full spectral-timing reprocessing model to simulate the
UV/optical lightcurves of NGC 5548. We show that disc reprocessing of
the observed hard X-ray lightcurve produces optical lightcurves with too much
fast variability as well as too short a lag time. 
Supressing the fast variability requires an intervening structure
preventing the hard X-rays from illuminating the disc. We propose this
is the disc itself, perhaps due to atomic processes in
the UV lifting the photosphere, increasing the scale-height, making it
less dense and less able to thermalise, so that it radiates low
temperature Comptonised emission as required to produce the soft X-ray
excess. The outer edge of the puffed-up Comptonised disc region emits FUV
flux, and can illuminate the outer thin blackbody disc but while this gives 
reprocessed variable emission which is much closer to
the observed UV and optical lightcurves, the light travel
lags are still too short to match the data. We reverse engineer a solution 
to match the observations and 
find that the luminosity and temperature of the lagged emission is
not consistent with material
at the light travel lag distance responding to the irradiating flux
(either FUV or X-ray) from the AGN. We conclude that the UV/optical
lags of NGC 5548 are not the light travel time from X-ray reprocessing, 
nor the light travel time from FUV reprocessing, but instead could be the 
timescale for the outer blackbody disc vertical structure to
respond to the changing FUV illumination. 

\end{abstract}


\section{Introduction} \label{sec:introduction}

The emission from Active Galactic Nuclei (AGN) is typically variable,
with faster variability seen at shorter wavelengths. This variability
can be used as a tool to probe the surrounding structures, with
reverberation mapping of the broad line region being an
established technique. However, the same techniques can be used to probe the
structure of the accretion flow itself. Hard X-ray illumination of the
disc should produce a lagged and smeared thermal reprocessing
signal. Larger radii in the disc produce lower temperature emission,
so this disc reprocessing picture predicts longer lags at longer
wavelengths. Such differential lags are now starting to be seen
(Sergeev et al. 2005; McHardy et al. 2014; Edelson et al. 2015; Fausnaugh
et al. 2015; McHardy et al. 2016) confirming qualitatively that we are
indeed seeing reprocessing from radially extended, optically thick
material, as expected from a disc.

However, quantitatively, the picture runs into difficulties. It has long
been known that the implied size-scales are larger by a factor of a
few compared to the expected sizes from a Shakura-Sunyaev disc
(e.g. Cackett, Horne \& Winkler 2007).  Independent size-scale estimates
from microlensing also imply that the optical/UV emission region is
larger than expected by a similar factor (e.g. Morgan et al. 2010). Yet
the optical/UV spectrum shows a strong rise to the blue, and can be
fairly well fit by the emission expected from the outer disc regions
in a Shakura-Sunyaev model (e.g. Jin et al. 2012;
Capellupo et al. 2015), though there are discrepancies in detail
(e.g. Davis, Woo \& Blaes 2007).

Thus qualitatively the disc reprocessing picture appears sound, yet
quantitatively it fails to match the data. This is perhaps not
surprising for several reasons. Irradiation can change the structure
of the disc e.g. by flaring it, as well as by changing the local
heating (Cunningham 1976). Secondly, the spectra of AGN are
clearly not simply a disc.  The hard X-ray corona itself must be
powered by accretion, pointing to a change from a pure Shakura-Sunyaev
disc structure in the inner regions (e.g. Done et al. 2012). There is
also the generic additional component seen in AGN, the soft X-ray
excess, which again points to some change in disc structure which is
not captured by the simple Shakura-Sunyaev equations (e.g. Gierlinski
\& Done 2004; Porquet et al. 2004).

Here we use the unprecedented SWIFT and HST lightcurves collected by
the 2014 campaign on NGC 5548, which spans 120 days with sampling of
0.5 days, across 9 continuum bands from V to hard X-rays (Edelson et
al.  2015). On long (month-year) timescales, the optical
and X-ray lightcurves are well correlated, but the
optical lightcurves show more variability than the hard X-rays, 
ruling out the simplest reprocessing models as accounting for all the 
optical variability in this source (Uttley et
al. 2003, see also Arevalo et al 2008; 2009). 
However on day timescales the optical lightcurves lag
behind the X-rays, with lag times increasing with wavelength as
expected of disc reprocessing, and this is sampled in unprecedented
detail by the 2014 campaign lightcurves. We use these to
quantitatively test disc reprocessing models using a full model of
illumination and reprocessing, with the aim to reproduce both the
spectrum and variability of the source.

We first examine traditional disc reprocessing models and confirm that
these cannot explain the lag timescales of the optical/UV lightcurves
of NGC 5548.  However, these quantitative models reveal another more
fundamental conflict, which is that the UV and optical lightcurves
cannot be produced by reprocessing of the observed hard X-ray flux.
Disc reprocessing smears the variability by a similar timescale to the
lag. The UV and optical are lagged by 1-2 days, but are much smoother
than the hard X-ray lightcurve, with no sign of the rapid 1-2 day
variability seen in the X-rays (see also Arevalo et al. 2008 and the discussion of variability in Lawrence 2012). This
clearly shows that the fast hard X-ray variability is not seen by the outer disc, 
so not only are the observed X-rays not the driver for reprocessing, 
but the outer disc must be shielded from the
observed X-rays. Instead, the FUV (represented by the HST lightcurve) is a
much better match to the observed smoothness of the optical/UV
lightcurves of NGC 5548.  We incorporate these two aspects together in
a model where the soft X-ray excess is produced from the inner regions
of a moderately thickened disc which emits optically thick Compton (hence we name it the Comptonised disc) and which shields the outer blackbody (BB) disc from
direct hard X-ray illumination from the central corona. 
The observed difference between the FUV
and soft X-ray lightcurves clearly shows that this is not a single
component, so we assume that the outer regions of the thickened Comptonised disc
structure produce the FUV which can illuminate the outer thin BB
disc. However, this still predicts light travel time lags which are shorter than
observed. Increased flaring of the outer BB disc does not help because
these large radii regions with the required long lag times are too
cool to contribute significant optical flux due to their large area.

We explore the suggestion that the longer than expected lags come from
the contribution of the classic BLR (H$\beta$ line) to the optical and
UV emission (Korista \& Goad 2001), but this does not work either as these do
not contribute enough lagged flux.
Since all known models fail, we reverse engineer a geometry which
can fit both spectral and variability constraints. We use
the observed optical lags in the different wavelength bands to
constrain the luminosity and temperature of blackbody components at 
different lag times. We find the observations can be well matched by a single blackbody
component lagged by 6 days behind the FUV irradiation, consistent with reprocessing on a population of clouds interior to the classic BLR as suggested by Lawrence (2012). However, the
derived area of the reprocessor at this light travel time distance is far too small to
intercept enough of the AGN luminosity (either FUV, X-ray or total) to
give the observed luminosity of the lagged component.  We conclude that the
reprocessing timescale is not set by the light travel time.
Interestingly, the temperature of the required lagged 
component is close to $10^4$~K, which is the trigger for the onset of
the dramatic disc instability connected to hydrogen ionisation (see e.g. Lasota 2001)
so it could instead be linked to the changing structure of the disc at this point.

Fundamentally, the
time lags give us the wrong answers because the reverberation signal
is not from a thin BB disc responding on the light travel time to illumination
of either X-rays or FUV. We suggest it is instead from the inner edge of the thin BB disc
changing its structure in response to an increase in FUV illumination and 
expanding on the vertical timescale to join the larger scale height Comptonised disc region.

\section{Energetics of Disk Illumination and Reprocessing}

For all models we fix $M=3.2\times 10^7M_\odot$ (as used by Edelson et
al. 2015, from Pancoast et al. 2014: see also their Erratum 2015). This
is similar to the Denny et al. (2010) estimate of $4.4\times
10^7M_\odot$, though a factor of $\sim 2$ smaller than the Bentz et al. (2010)
estimate of $7.8_{-2.7}^{+1.9}\times 10^7M_\odot$. We also fix
distance $D=75$\,Mpc, and spin $a=0$, and assume an inclination angle of
$45^\circ$.

The unabsorbed, dereddened broadband spectrum of NGC 5548 from Summer
2013 is shown in Mehdipour et al. (2015). The
X-ray flux is very hard, with photon index $\Gamma=1.6$, and the
X-rays dominate the energy output of the source, peaking in $\nu F_\nu$ at
$\sim 8\times 10^{-11}$~ergs~cm$^{-2}$~s$^{-1}$ at 100~keV.  The
simultaneous optical/UV spectrum looks similar to that expected from
an outer thin BB disc (though its shape is subtly different: Mehdipour et
al. 2015). The flux in UVW1 (at around 5~eV) has $\nu F_\nu\approx 5\times
10^{-11}$~ergs~cm$^{-2}$~s$^{-1}$.

We use the {\sc optxagnf} model in {\sc xspec} to find reasonable
physical parameters for the accretion flow.  This assumes that the
mass accretion rate is constant with radius, with fully relativistic
Novikov-Thorne emissivity per unit area, $L_{NT}(r)$, 
with dimensionless radius $r=R/R_g$ for $R_g=GM/c^2$, but that this
energy is dissipated in a (colour temperature corrected) BB disc only
down to some radius $r_{cor}$, with the remainder split between
powering an optically thick Compton component (which provides the soft X-ray excess emission) and an optically thin Compton component, which models the hard X-ray coronal
emission. This code calculates the angle averaged spectrum, so we
boost the normalisation by a factor $\cos i/\cos 60=1.41$ to roughly
account for our assumed inclination angle. 
If we assume that all the power within
$r_{cor}=70$ goes to make the hard X-ray corona ($f_{cor}=1$), we find that we can match
the UVW1 and X-ray flux for $\log L/L_{Edd}=-1.4$ with $r_{cor}=70$.

This does not necessarily mean that the geometrically
thin BB disk itself is not present below $70\,R_g$, only that the accretion power is not
dissipated within this structure (Svensson \& Zdziarski 1994; Petrucci
et al 2010).  However, there is additional information in the hard
X-ray spectrum which does point to this conclusion.  An optically
thick BB disc cannot be present underneath an isotropically emitting
corona in this object, as such a disc will intercept around half of
the hard X-ray flux, giving a strong Compton hump which is not present
in the NuSTAR data (see spectrum in Mehdipour et al. 2015 and Ursini et
al. 2015). Thermalisation of the non-reflected emission also produces
too many seed photons for the X-ray source to remain hard (Haardt \&
Maraschi 1991; 1993; Stern et al. 1995; Malzac et al. 2005; Petrucci et
al. 2013).  Together, these imply that either the X-ray source is very
anisotropic, or the disc truly truncates for sources with hard X-ray
spectra.

Black hole binaries similarly show hard X-ray spectra and small
Compton hump in their low/hard state, but here there are clear limits
on the possible anisotropy from comparing sources with different
binary inclination angles (Heil, Uttley \& Klein-Wolt 2015).
This argues strongly for true truncation of the BB accretion disc, as
does the currently popular Lense-Thirring precession model for the
origin of the low frequency QPO's seen in high inclination binary
systems (Ingram, Done \& Fragile 2009). Hence we
assume the BB disc is truly truncated in these low luminosity AGN
(see also Petrucci et al. 2013; Noda 2016).

The irradiation pattern strongly depends on the relative geometry of
the hard X-ray source and BB disc. In the black hole binaries, there is
evidence from the complex pattern of energy dependent lags that the
hard X-ray source is somewhat radially extended (Kotov, Churazov \&
Gilfanov 2001; Ingram \& Done 2012), as is also expected if it is some
form of hot, radiatively inefficient accretion flow such as an
advection dominated accretion flow (Narayan \& Yi 1995). We assume
that the extended hard X-ray source has volume emissivity $\propto
L_{NT}(r)/r$ and neglect light-bending and red/blueshifts to work out
the irradiating flux as in the appendix of Zycki et al (1999), i.e.
\begin{equation}
F_{rep}(r) = \frac{f_{irr}L_{cor}\cos(n)}{4\pi (\ell R_g)^2}
\end{equation}
where $\ell$ is the distance from the hard X-ray source to the disc
surface element, $n$ is the angle between the source and normal to the
disc, and $f_{irr}$ is the fraction of coronal hard X-ray luminosity 
($L_{cor}$) which thermalises. We set $f_{irr}=1$ in this section in 
order to see the maximum irradiation flux. 
We then assume that
\begin{equation}
T_{eff}(r) = T_{grav}(r) \left(\frac{F_{rep}(r)+F_{grav}(r)}{F_{grav}(r)}\right)^{1/4}
\end{equation}
and model the resulting optically thick disc emission at this radius
as a (colour temperature corrected) blackbody. However, the colour
temperature correction makes very little difference for the black hole
mass, mass accretion rate and $r_{cor}$ used here, as the disc peaks
at 5.5~eV ($\sim 2500$\AA) so is too cool for the colour temperature
correction to be significant.

\begin{figure} 
\centering
\begin{tabular}{l}
\leavevmode  
\includegraphics[width=8cm]{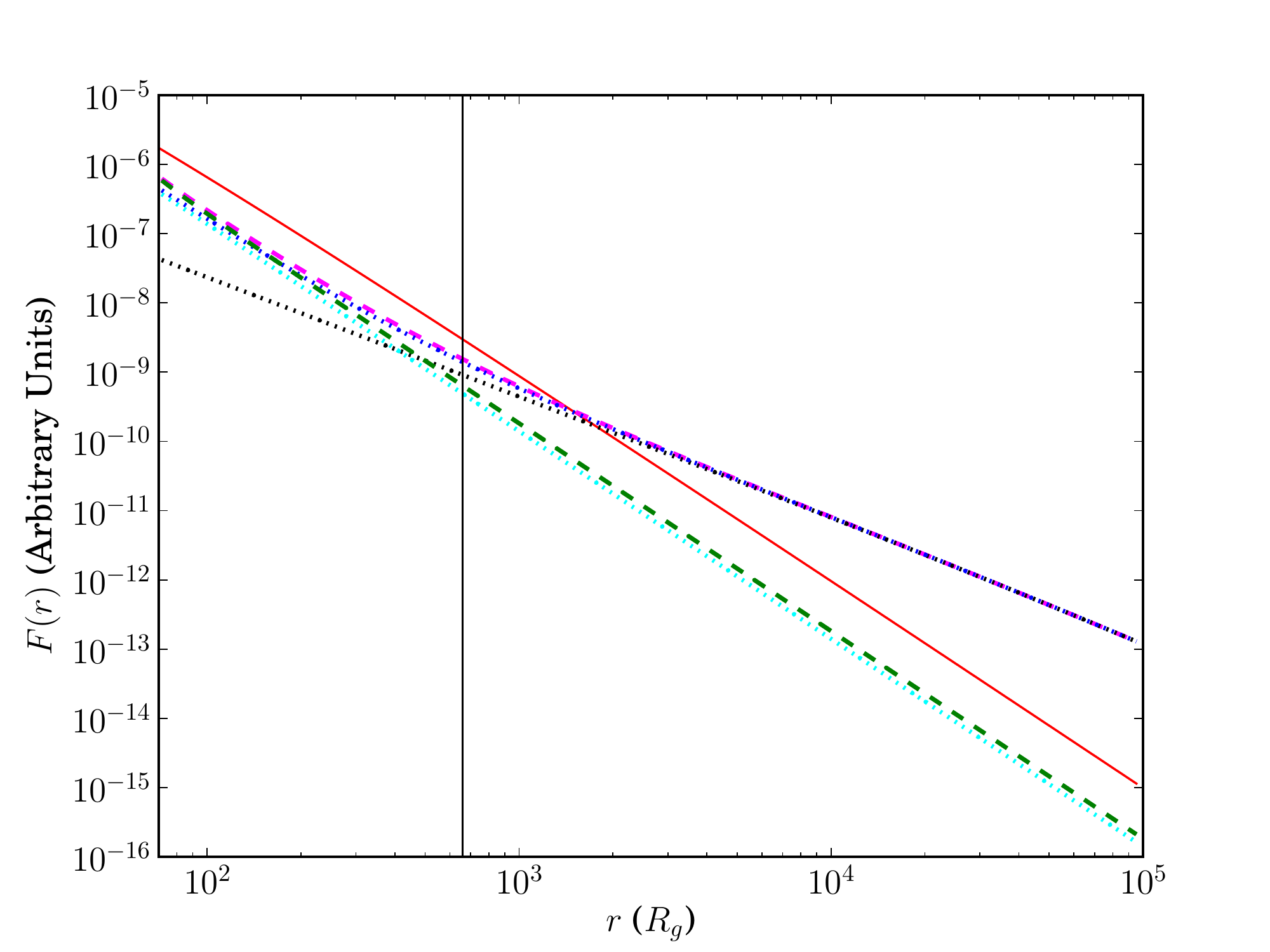} \\
\end{tabular}
\caption{Illuminating flux as a function of BB disc radius. Green and magenta dashed lines show flat ($h=0.1$) and flared ($h=h_0 (r/r_0)^{9/7}$, where $h/r=0.1$ at $r=660$) discs, respectively, irradiated by an extended spherical source with $r=70$. Cyan and blue dotted lines show same flat disc and flared discs, respectively, illuminated by a central point source with height $h_x=10$. Black dotted line shows flared disc illuminated by a central point source with $h_x=0$. Red solid line shows gravitational flux dissipation of BB disc. Black solid line shows self-gravity radius.}
\label{fig0}
\end{figure}

We first assume the extended X-ray source illuminates a flat BB disc (formally, we give this disc constant height of
$h=H/R_g=0.1$). The green dashed line in Fig.\ref{fig0} shows the resulting
irradiation flux per unit area on the equatorial plane. This is
$\propto r^{-3}$ at large $r$, similar to the intrinsic
gravitational flux dissipation (red), and is around 10-20\%
of the intrinsic flux at each radius.  We compare this to the
mathematically simpler form of a lamppost at height $h_x=10$ (cyan dotted line,
where $h=H/R_g$) illuminating the BB disc, and show that the two are
comparable at all disc radii. This is important as it shows that using
lamppost illumination is not necessarily the same as assuming that the
source is a lamppost (compact source on the spin axis). The more
physical extended source geometry is identical in its illumination
properties to the mathematically simpler lamppost.

It is unlikely that the BB disc remains flat under illumination. Cunningham
(1976) shows that the disc structure responds to illumination, and can
form a flared disc with height $h=h_0 (r/r_0)^{9/7}$, where $h_0$ is
the disc height at its outer radius $r_0$.  We show the resulting
illumination for an extended X-ray source (Fig.\ref{fig0}, magenta dashed line), and a lamppost source of height
$h_x=10$ (blue dotted line) and a central source ($h_x=0$: black dotted line) for a flared disc with $h/r=0.1$ at
$r=660$. Clearly, irradiation can dominate over gravitational energy release for such a flared disc, but
only at large radii. 

Crucially, irradiation changes the predicted lag-wavelength profile of
the BB disc. The standard argument for a lag time $\tau\propto
\lambda^{4/3}$ comes from assuming that the wavelength at which the
disc peaks at each radius is $\lambda_{max}\propto 1/T\propto
L_{NT}(r)^{-1/4}\propto (M\dot{M})^{-1/4} R^{3/4}$. Hence
$R\propto\tau\propto (M\dot{M})^{1/3}\lambda_{max}^{4/3}$. However, in
the irradiation dominated region, the emissivity $L(r)\propto
r^{-12/7}$, so the lags are no longer expected to go as $\tau\propto
\lambda^{4/3}$ but as $\propto \lambda^{7/3}$. 

Fig.\ref{fig0} shows irradiation
only gives $L(r)$ which is substantially different to $r^{-3}$ at
$r>2000$, so all our irradiation models predict $\tau\propto
\lambda_{max}^{4/3}$ for $r<1000$.  The self-gravity radius for a disc
with these parameters is only 660\,$R_g$ (black vertical line: Laor \&
Netzer 1989, using Shakura-Sunyaev $\alpha=0.1$), so that truncating the BB disc
at this point means that irradiation never dominates, and there
is only a factor of $\sim 2$ between the flared and flat disc
illumination fluxes at $r=660$. 

In the following reprocessing models we wish to maximise irradiation. We therefore use the lamppost at $h_x=10$ illuminating a
flared BB disc (Fig.\ref{fig0}, blue dotted line; illumination pattern identical to an extended source spherical source with $r=70$) in all subsequent models.

\subsection{Calculation of Cross Correlation Functions}

Throughout the paper we compare lags between lightcurves by
calculating the cross correlation function (CCF). For
two lightcurves $x(t)$ and $y(t)$, which are evenly sampled on time
$\Delta t$ so $t=t_0+i\Delta t$, the CCF as a function of lag time
$\tau=j\Delta t$ is defined as:

\begin{equation}
CCF(\tau) = \frac{\sum (x(i)-\bar{x})(y(i-j)-\bar{y})}
{[(\sigma_x^2-\sigma_{ex}^2) (\sigma_y^2-\sigma_{ey}^2)]^{1/2} }
\end{equation}

\noindent where the sum is over all data which contribute to the lag
measurement, so there are a smaller number of points for longer
lags. The averages, ($\bar{x},\bar{y}$), and measured variances
($\sigma^2_x,\sigma^2_y$), and error bar variances
($\sigma^2_{ex},\sigma^2_{ey}$), are also recalculated for each $\tau$
over the range of data used. With this definition, then $CCF(\tau)=1$
implies complete correlation.

\begin{figure} 
\centering
\begin{tabular}{l}
\leavevmode  
\includegraphics[width=8cm]{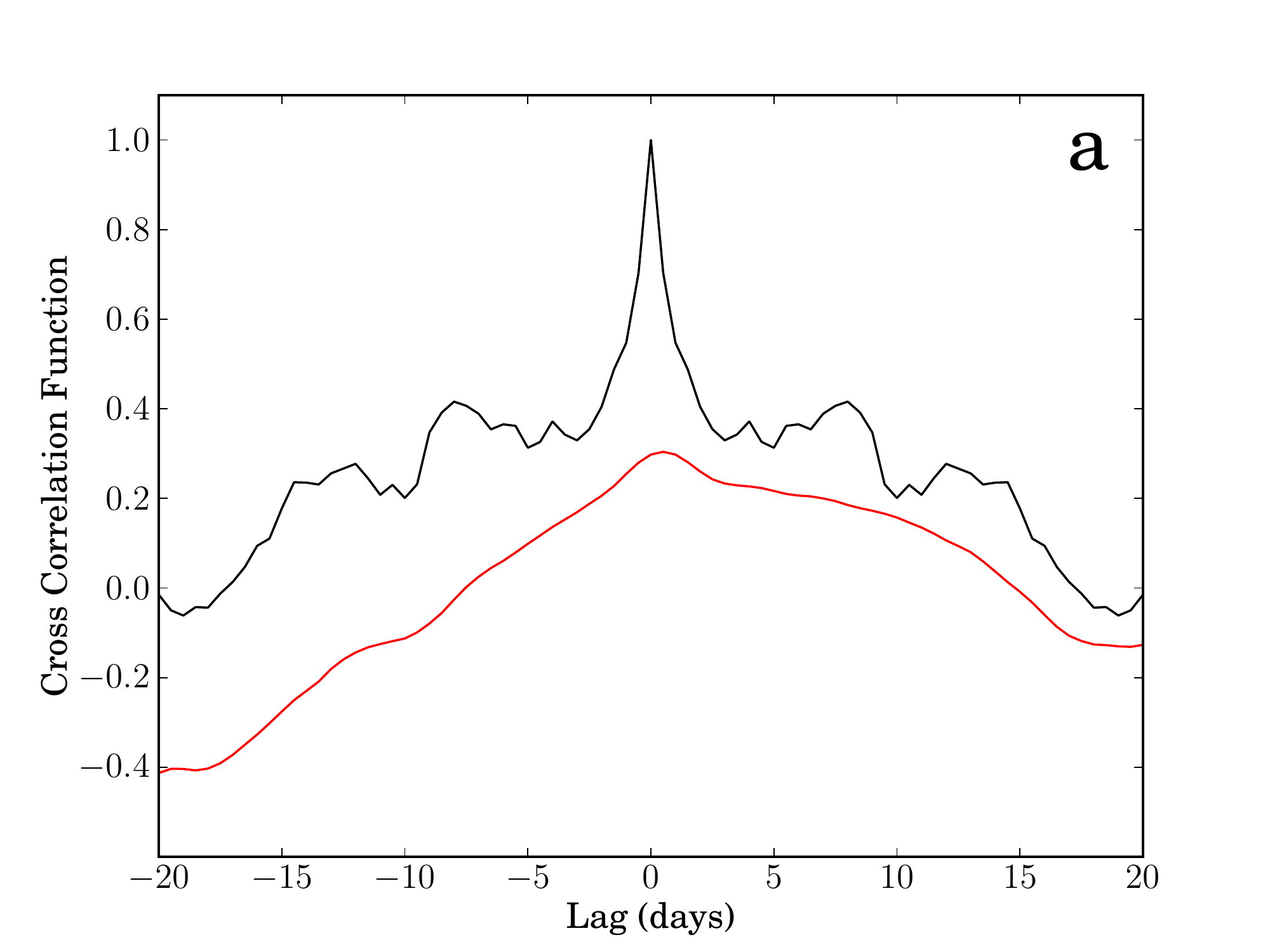} \\
\includegraphics[width=8cm]{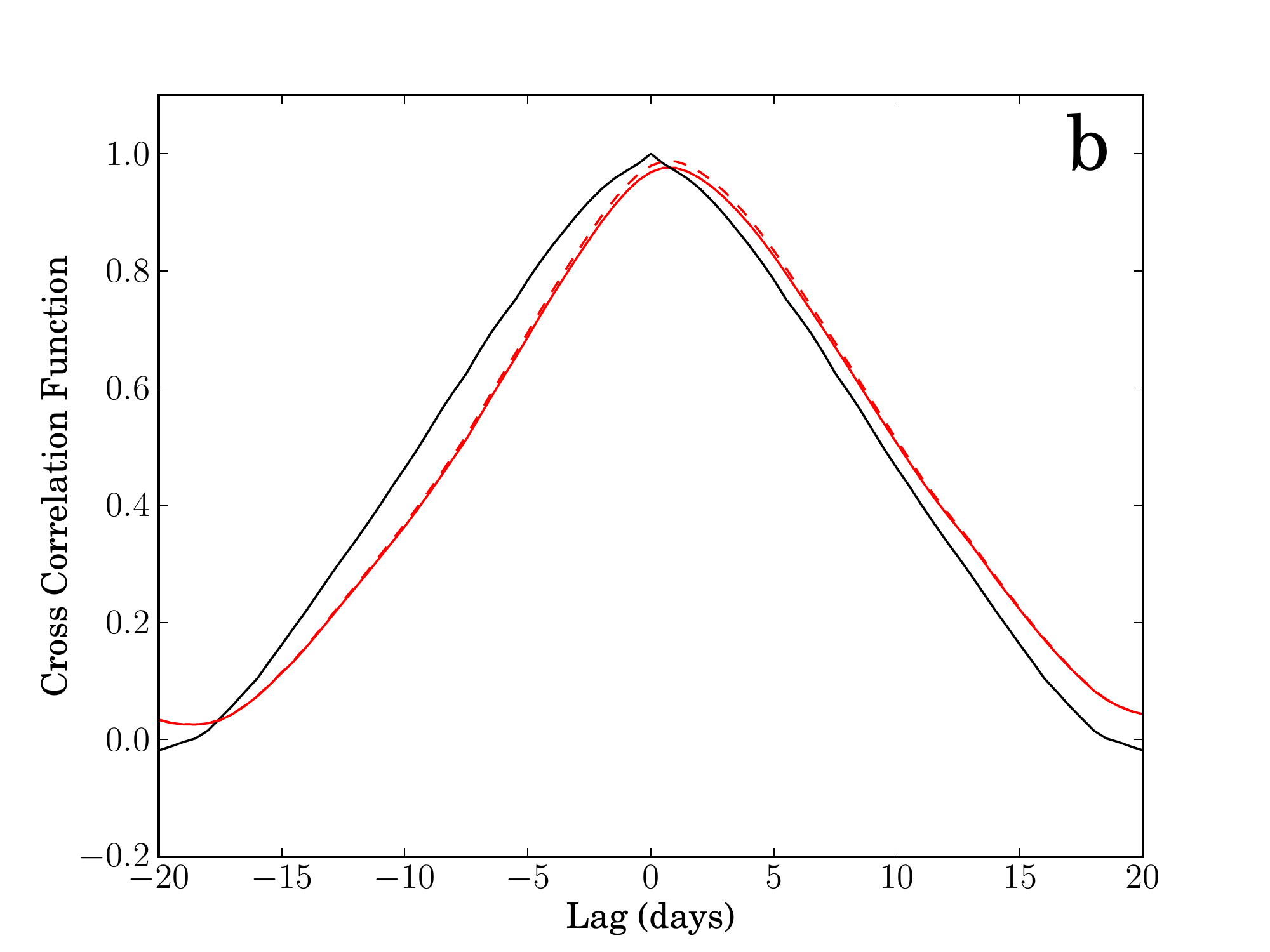} \\
\includegraphics[width=8cm]{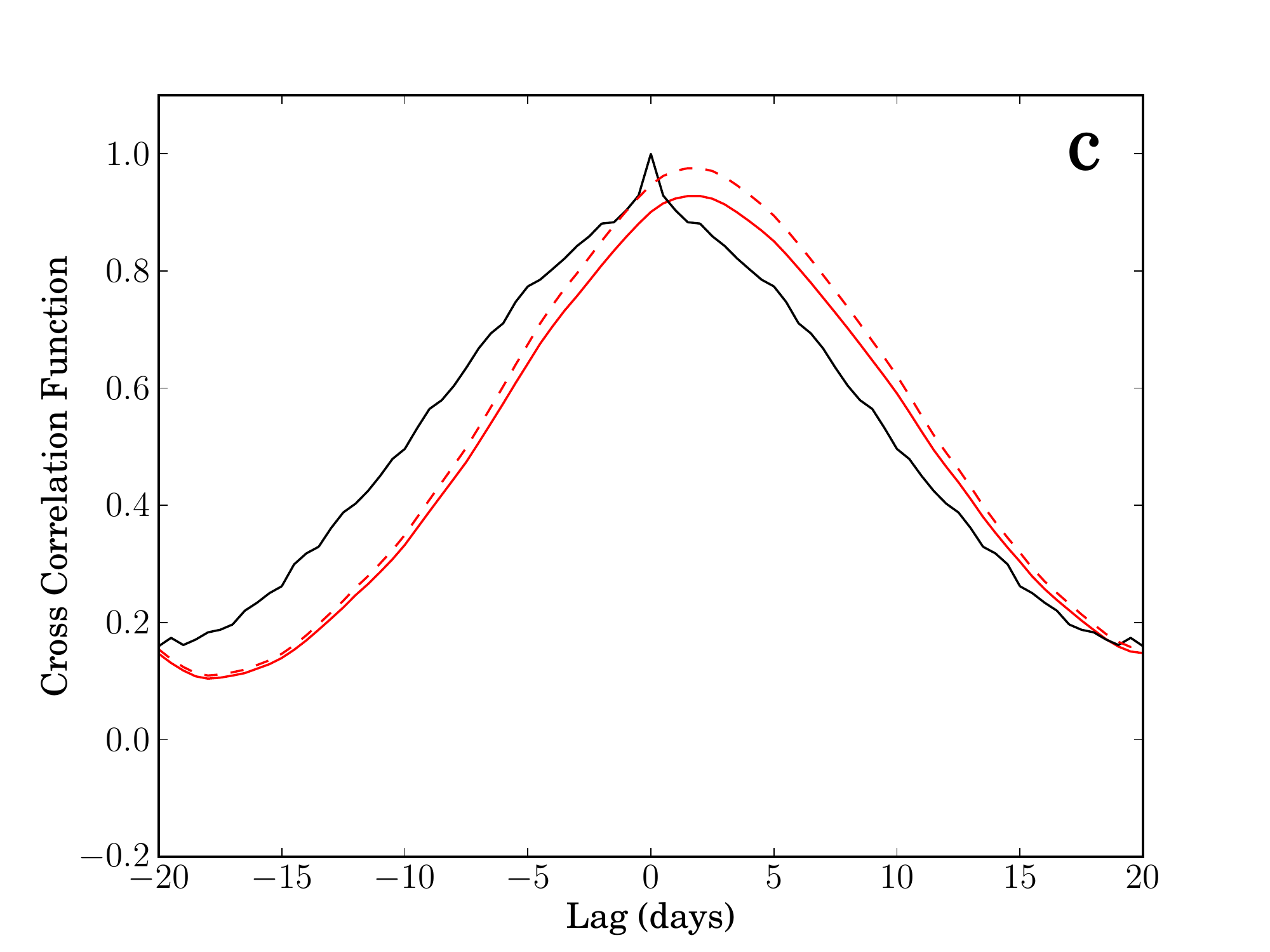} \\
\end{tabular}
\caption{Solid red lines show cross correlation functions calculated from the observed lightcurves of NGC 5548: (a) FUV with respect to hard X-rays. (b) UVW1 with respect to FUV. (c) V band with respect to FUV. Solid black lines in a, b \& c show the autocorrelation functions of the hard X-ray, UVW1 and V bands, respectively. The narrow peaks at zero lag in b \& c are due to correlated errors introduced by interpolating the lightcurves. Red dashed lines in b \& c show the cross correlation functions (UVW1 w.r.t. FUV and V band w.r.t. FUV, respectively) after correcting for these correlated errors.}
\label{figccfs}
\end{figure}

However, real data are not exactly evenly sampled. Interpolation is
often used to correct for this, but this  can be done in multiple ways
(Gaskell \& Peterson 1987).  The red solid lines in Fig.\ref{figccfs}
shows the interpolated CCF, computed by linearly interpolating both
lightcurves onto a grid of $0.1$\,d spacing and then resampling to
produce evenly sampled lightcurves with $dt=0.5$\,d. This
interpolation scheme introduces correlated errors, and it is not
simple to correct for these so we first set
$\sigma_{ex}=\sigma_{ey}=0$. The correlation is very poor between the hard
X-ray and FUV lightcurve (Fig.\ref{figccfs}a), while the FUV and UVW1
are consistent with almost perfect correlation with a lag of $\sim 0.5$~days
(Fig.\ref{figccfs}b). The correlation without considering the error
bar variance is slightly worse between  FUV and V band, while the lag is
somewhat longer (Fig.\ref{figccfs}c). 

The neglect of the error bar variance can suppress the correlation, so
we first investigate how much of the lack of correlation in
Fig.\ref{figccfs}a-c is due to this. We assess the size of this
effect by calculating the autocorrelation function (ACF). For an
evenly sampled lightcurve with independent errors, the error bars are
correlated only at zero lag, giving an additional spike at zero on top
of the intrinsic ACF shape. The interpolated lightcurves have
correlated errors on timescales of the interpolation, so instead of a
spike at zero lag, these form a component with width $\sim 0.5~d$ on
top of the intrinsic ACF.  We show the ACF of the interpolated 
UVW1 and V band lightcurves as the solid black lines in
Fig.\ref{figccfs}b\&c, respectively. We do not show the ACF of the FUV
lightcurve as this has such small errors and such good sampling that
there is negligible error bar variance in this lightcurve.  We fit the
ACFs with two Gaussians, one broad to model the intrinsic ACF, and one
narrow to represent the correlated errors, both of which should peak
at zero lag. This gives the value of $\sigma_e^2$ that must be
subtracted in order for the broad Gaussian to peak at unity. 
For UVW1 we find that $\sigma_e^2 = 0.022
\sigma_{UVW1}^2$, and for the V band we find $\sigma_e^2 = 0.095 \sigma^2_V$.
We subtract this correlated error variance to get the error corrected
interpolated correlation function (Fig.\ref{figccfs}b\&c, dashed red
lines). The effect is quite small for the (already very good) FUV--UVW1
correlation, but makes a noticeable difference to the FUV--V band
cross-correlation. The intrinsic variability in the V band lightcurve
is then consistent with an almost perfect correlation with the
variable FUV lightcurve, but with a $\sim 2$ day lag (see also McHardy
et al. 2014; Edelson et al. 2015; Fausnaugh et al. 2015). 

In Fig.\ref{figccfs}a we show the hard X-ray ACF (black line). It is not possible to 
apply the Gaussian fitting method to the hard X-ray ACF as the fast varying 
hard X-rays show an intrinsic peak of correlated variability on $\sim0.5$\,d timescales (Noda 2016) 
that cannot easily be separated from the effects of any correlated errors. 
However, we calculate the error bar variance of the hard X-ray lightcurve from the data and find 
it is roughly $\sigma^2_e\sim0.013\sigma^2_X$. This is negligible and cannot 
explain the poor correlation shown by the hard X-ray--FUV CCF, indicating that the poor
correlation between the hard X-ray and FUV lightcurves is intrinsic to the process. 

\section{Lightcurves and Spectra from Blackbody Disc Reprocessing Hard X-ray Emission}

Not all the irradiating flux will thermalize, as some part will be
reflected. The reflection albedo depends on the ionization state of
the disc, but for such a hard spectrum it varies only from 
$0.3$ (neutral) to $0.5$ (completely ionised), giving
$f_{irr}=0.7-0.5$. We choose $f_{irr}=0.5$ in all subsequent models.

\begin{figure*} 
\centering
\begin{tabular}{l|r}
\leavevmode
\includegraphics[width=8cm]{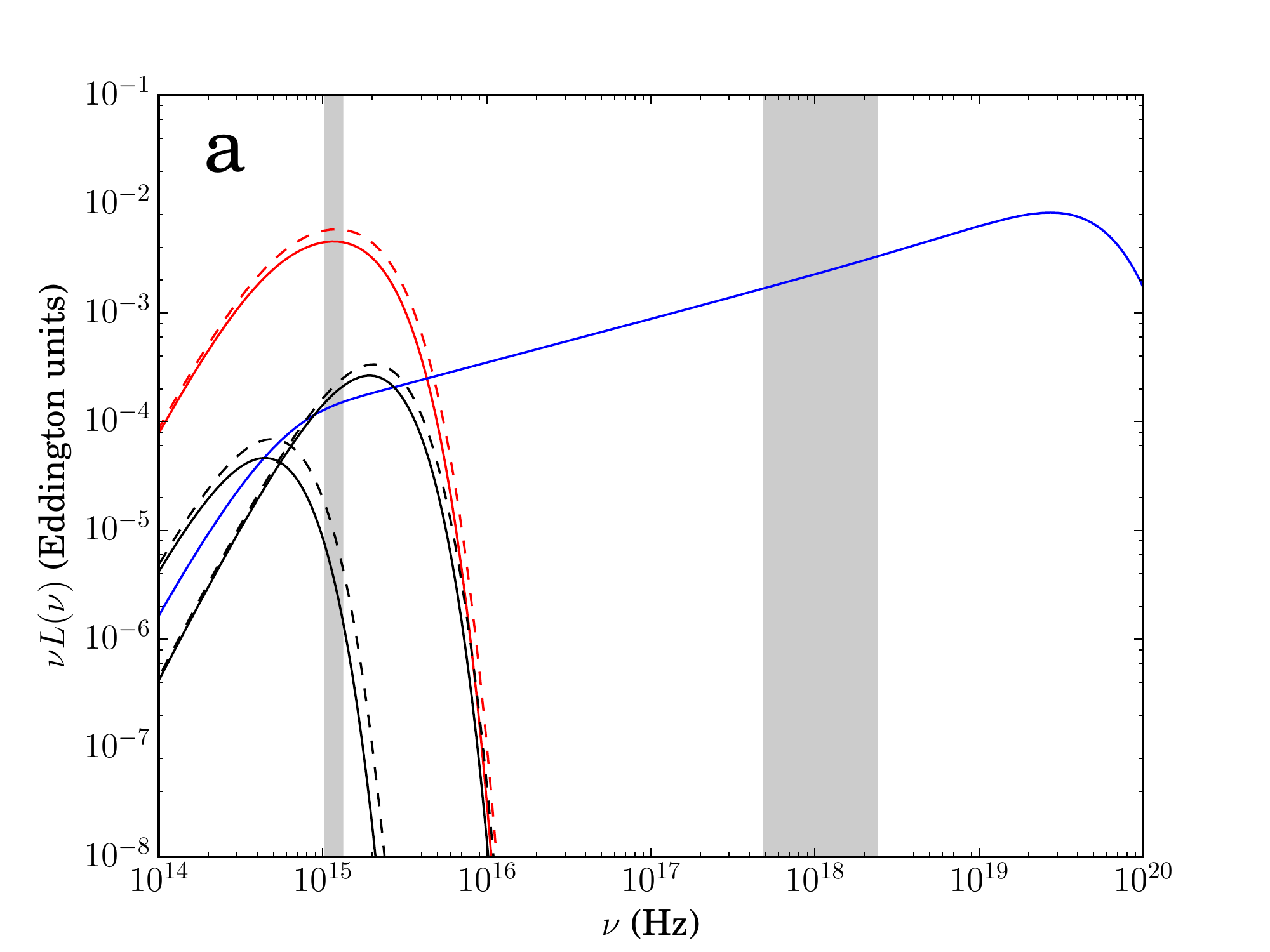} &
\includegraphics[width=8cm]{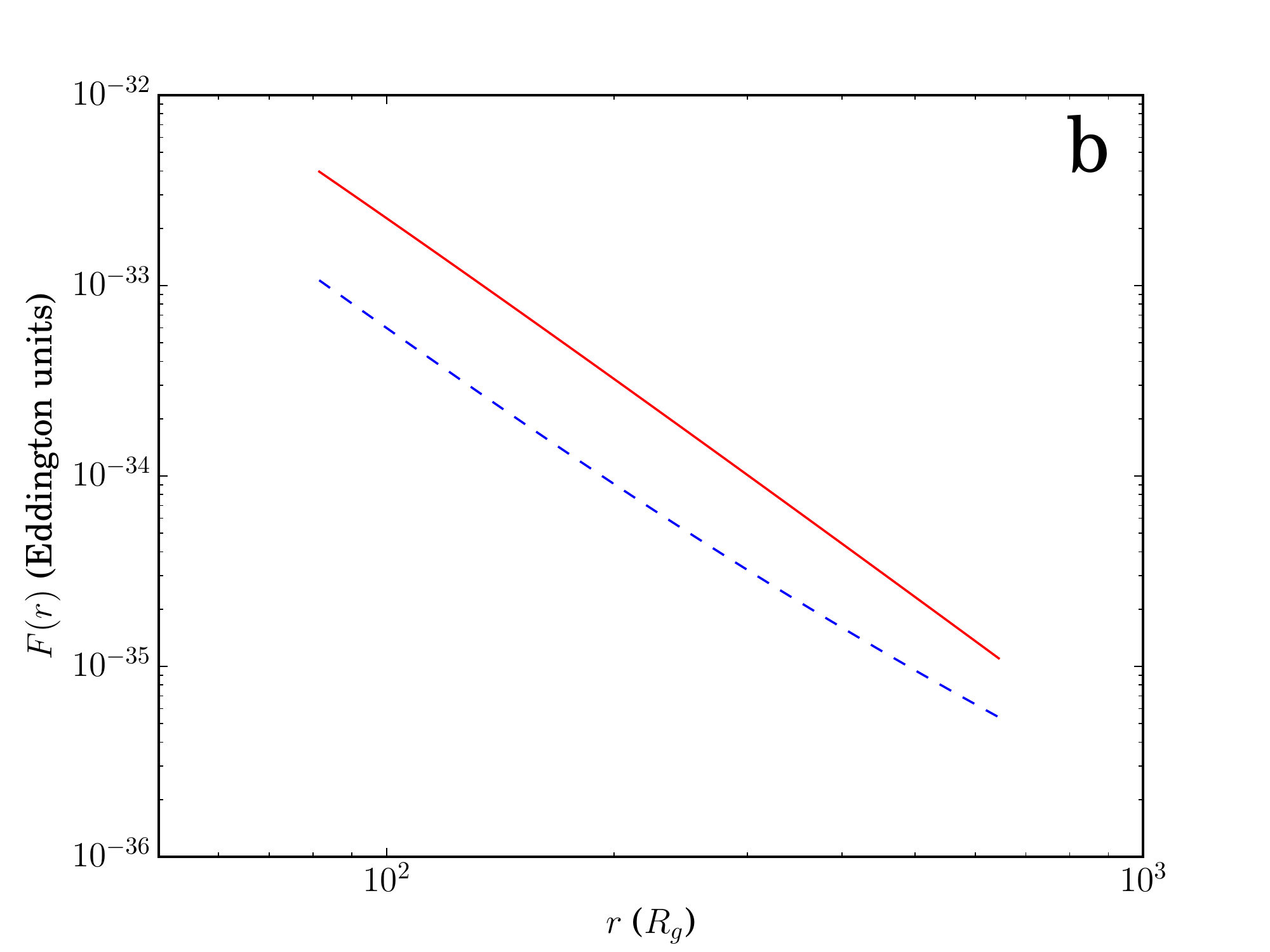} \\
\includegraphics[width=8cm]{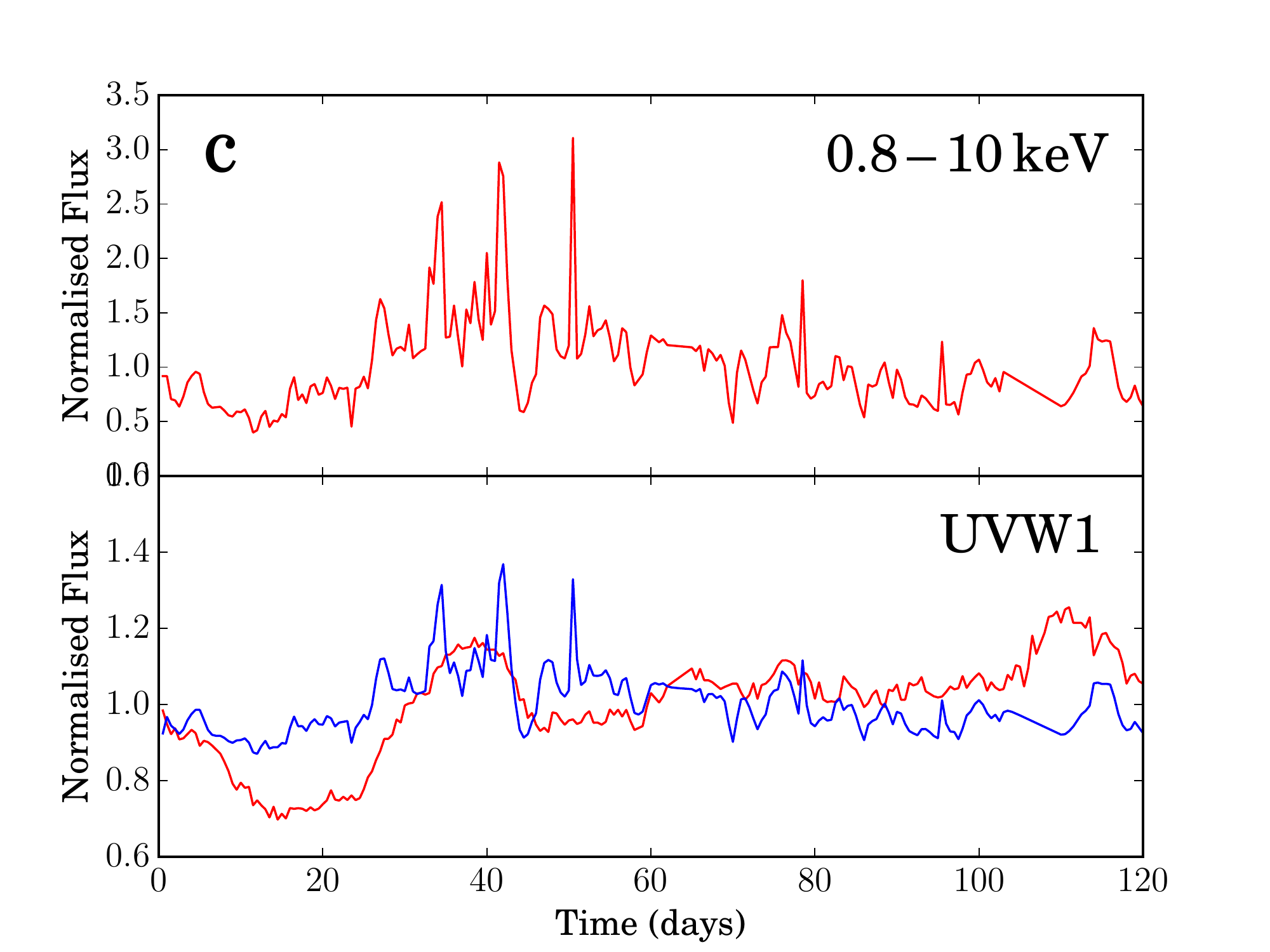} &
\includegraphics[width=8cm]{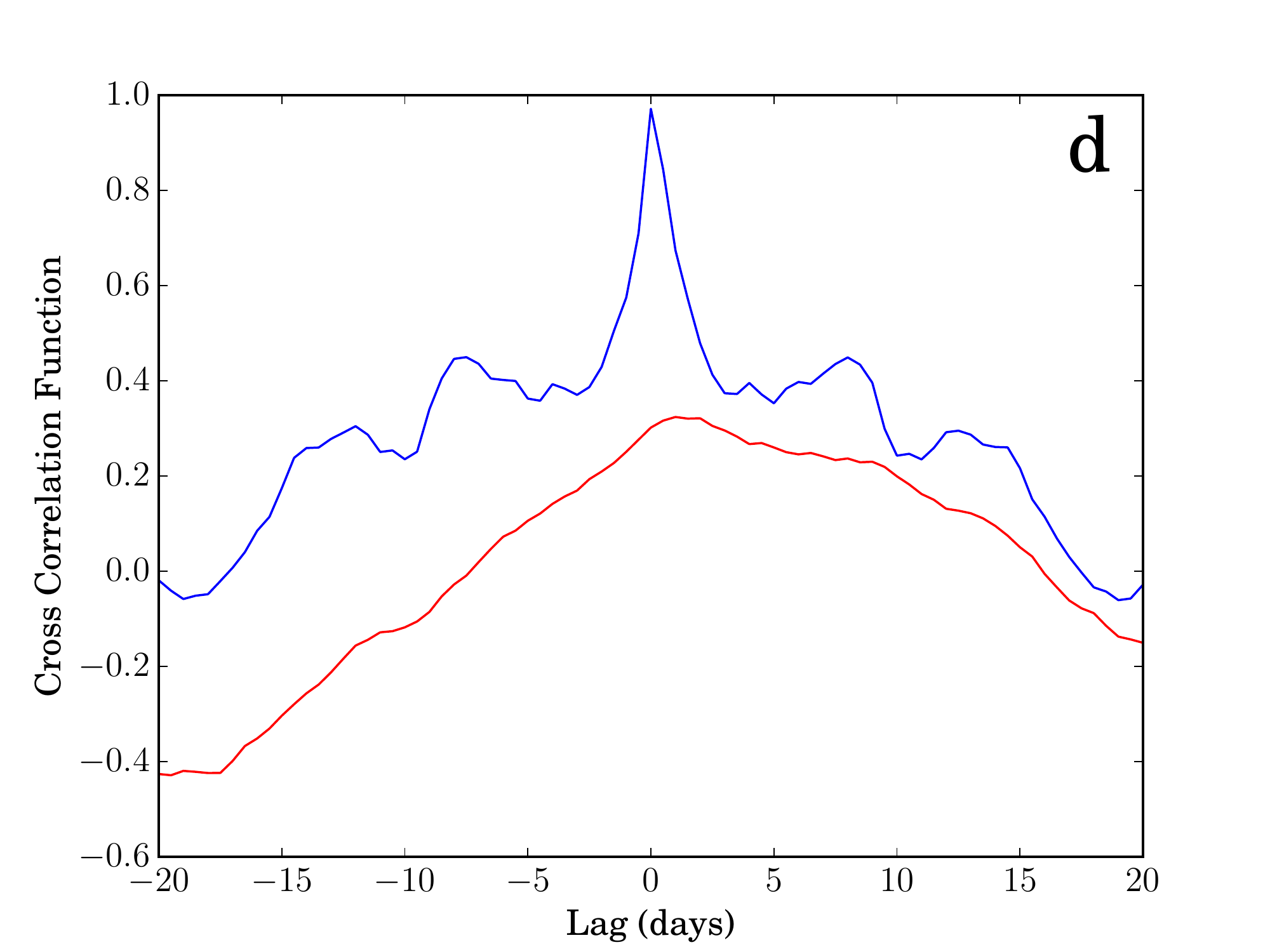} \\
\end{tabular}
\caption{Standard BB disc with $r_{cor}=70$ reprocessing hard X-ray emission. (a) Model spectrum: solid red line shows total intrinsic BB disc emission; dashed red line shows total disc emission including reprocessing; black lines show emission from outermost and innermost disc radii, with solid and dashed lines for intrinsic and intrinsic plus reprocessed emission, respectively; blue line shows hard coronal power law; grey shaded regions, from left to right, show location of UVW1 band and hard X-ray band respectively. (b) Intrinsic BB disc flux as a function of radius (red line) compared with illuminating hard X-ray flux available for reprocessing (dashed blue line). (c) Top panel shows observed hard X-ray lightcurve of NGC 5548 input into the model; bottom panel shows observed simultaneous UVW1 lightcurve (red) compared with simulated UVW1 lightcurve (blue). (d) Cross correlation function of observed UVW1 lightcurve with respect to observed hard X-ray lightcurve (red), and simulated UVW1 lightcurve with respect to observed hard X-ray lightcurve (blue). Positive lag values indicate the UVW1 band lagging behind the hard X-rays.}
\label{fig1}
\end{figure*}

We first explore the scenario in Section 2, where a flared BB disc with
scale-height 0.1 on its outer edge extends inwards from the self
gravity radius at $r_{out}=660$ down to $70$, with the flow then
forming a hot corona whose illumination can be approximated by a point
source at height $h_x=10$ above the black hole.  The solid red line in
Fig.\ref{fig1}a shows the intrinsic BB disc emission that results from
gravitational heating alone. The dashed red line shows the total disc
emission including additional heating by the illuminating corona. For
illustrative purposes we also show the emission from two individual
annuli in black, with the lower energy example corresponding to the
emission from $r_{out}=660$ and the higher energy example corresponding to
the emission from the innermost disc radius at $70$. Again solid lines
show the intrinsic emission from gravitational heating alone and
dashed lines show the total emission including reprocessing.

\begin{figure*} 
\centering
\begin{tabular}{l|r}
\leavevmode  
\includegraphics[width=8cm]{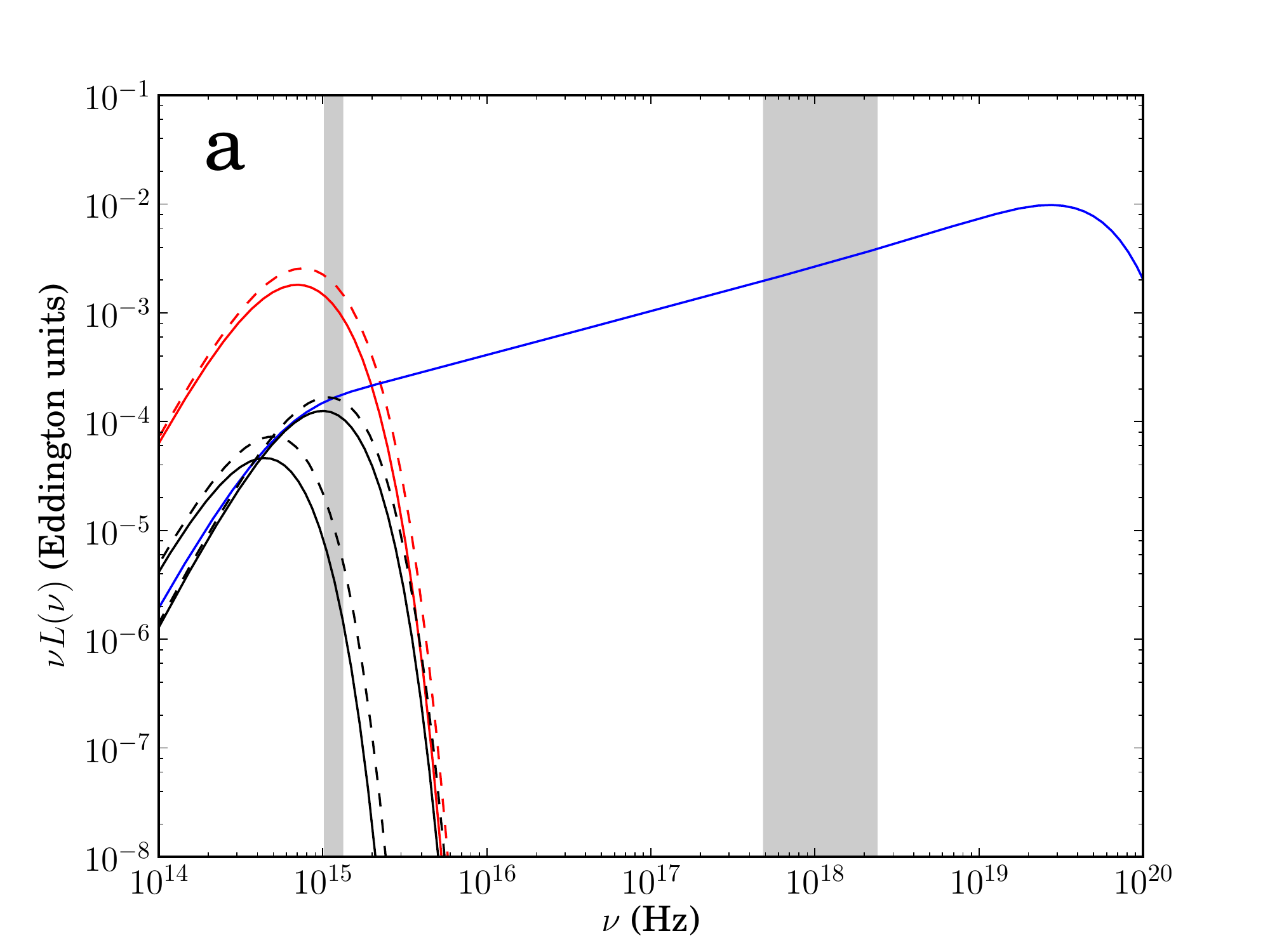} &
\includegraphics[width=8cm]{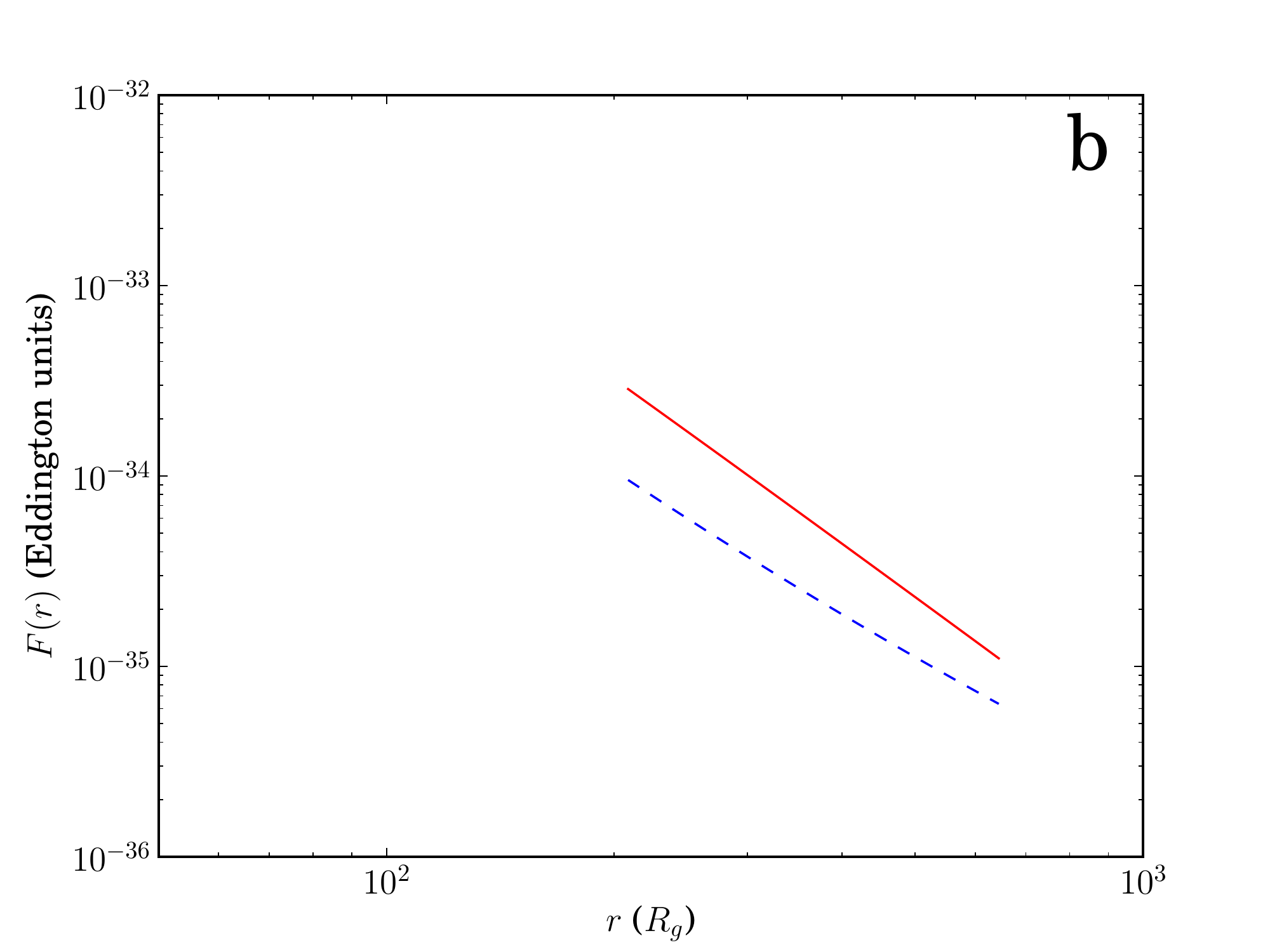} \\
\includegraphics[width=8cm]{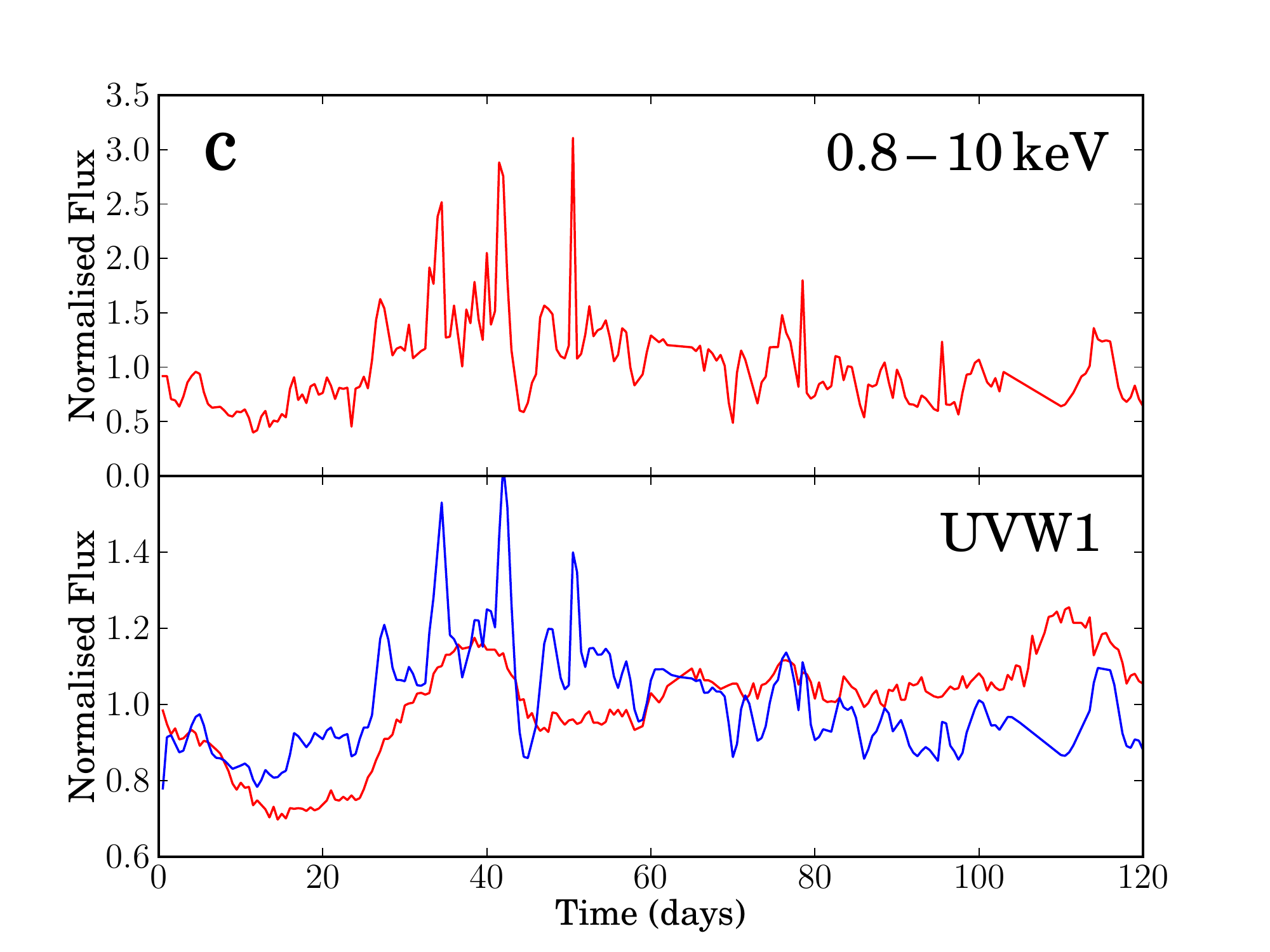} &
\includegraphics[width=8cm]{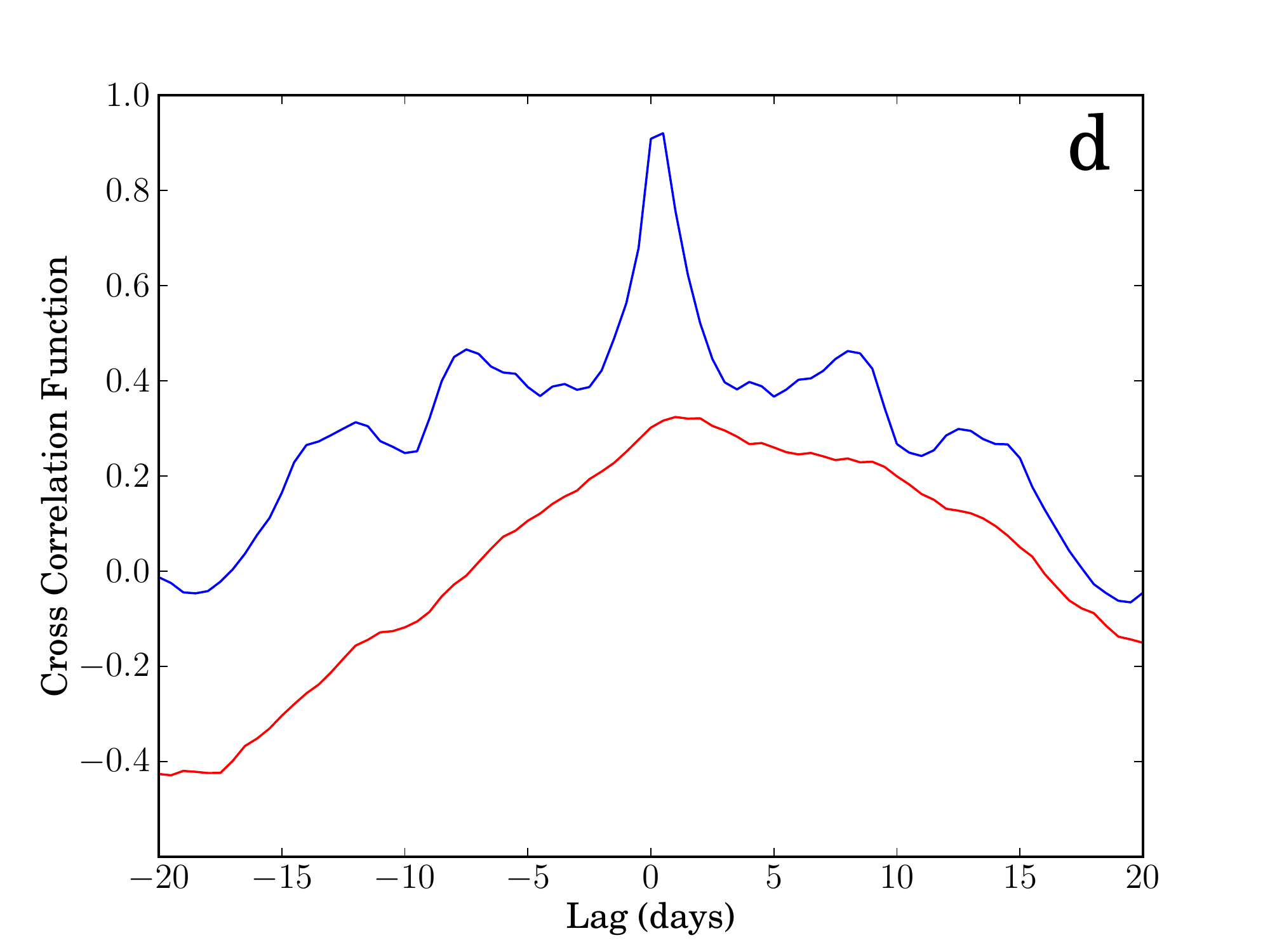} \\
\end{tabular}
\caption{As in Fig.\ref{fig1}, but now for a standard BB disc with $r_{cor}=200$ reprocessing hard X-ray emission.}
\label{fig2}
\end{figure*}

The dashed red line in Fig.\ref{fig1}a shows that reprocessing makes
very little difference to the total BB disc luminosity. This is because
the disc is dominated everywhere by the intrinsic emission (red solid
line in Fig.\ref{fig1}b) rather than by reprocessing (blue dashed line
in  Fig.\ref{fig1}b), and even increasing $f_{irr}$ to its maximum plausible value of $0.7$ cannot overcome this. Reprocessing does have slightly more effect at
larger radii (because the BB disc is flared), but the disc emission is dominated by the
smallest radii. 

So far we have considered the steady state or time-averaged
spectrum. In order to know how fluctuations in the hard X-ray flux
will produce changes in the UV/optical emission from the BB disc, we must
quantify how well each disc radius can respond to and reproduce
changes in the illuminating continuum. For each annulus in the disc we
calculate its transfer function following Welsh \& Horne (1991). This
accounts for light travel time distances to different radii within the
annulus and different azimuths within each radius. The transfer
function, $T(r,\tau)$, for a given radius, $r$, describes
what fraction of the reprocessed flux from that radius has a given time
delay, $\tau$, with respect to the illuminating continuum. The
fluctuations in the reprocessed flux from a given annulus is then:

\begin{equation}
F_{rep}(r,t) = \frac{f_{irr}\cos(n)}{4\pi (rR_g)^2}\int_{\tau_{min}}^{\tau_{max}} T(r,\tau) L_{cor}(t) \, d\tau
\end{equation}

This causes the effective temperature of the annulus to vary as:

\begin{equation}
T_{eff}(r,t) = T_{grav}(r) \left(\frac{F_{rep}(r,t)+F_{grav}(r)}{F_{grav}(r)}\right)^{1/4}
\end{equation}

The fluctuations in a given spectral band (e.g. UVW1) are then the sum
of the fluctuations in the emission from each annulus contributing
flux to that band. Fluctuations in reprocessed flux change the
relative contributions of individual annuli to the total band flux. An
increase in reprocessed flux increases the temperature of the
annulus. This both increases the luminosity of the annulus and shifts
the peak of its blackbody spectrum to higher energies. This may shift
the peak emission from smaller radii out of the bandpass and shift
more of the emission from larger radii, which usually peak below the
bandpass, to higher energies and so increase their contribution to the
total band flux. As such it is not appropriate to assume a band is
always dominated by emission from any one radius. By calculating the
effective temperature and corresponding blackbody spectrum from each
annulus at each timestep our code accounts for this. We neglect any
fluctuations in the intrinsic BB disc emission so that the only source of
disc variability is the reprocessed fluctuations.

\begin{figure*} 
\centering
\begin{tabular}{l|r}
\leavevmode  
\includegraphics[width=8cm]{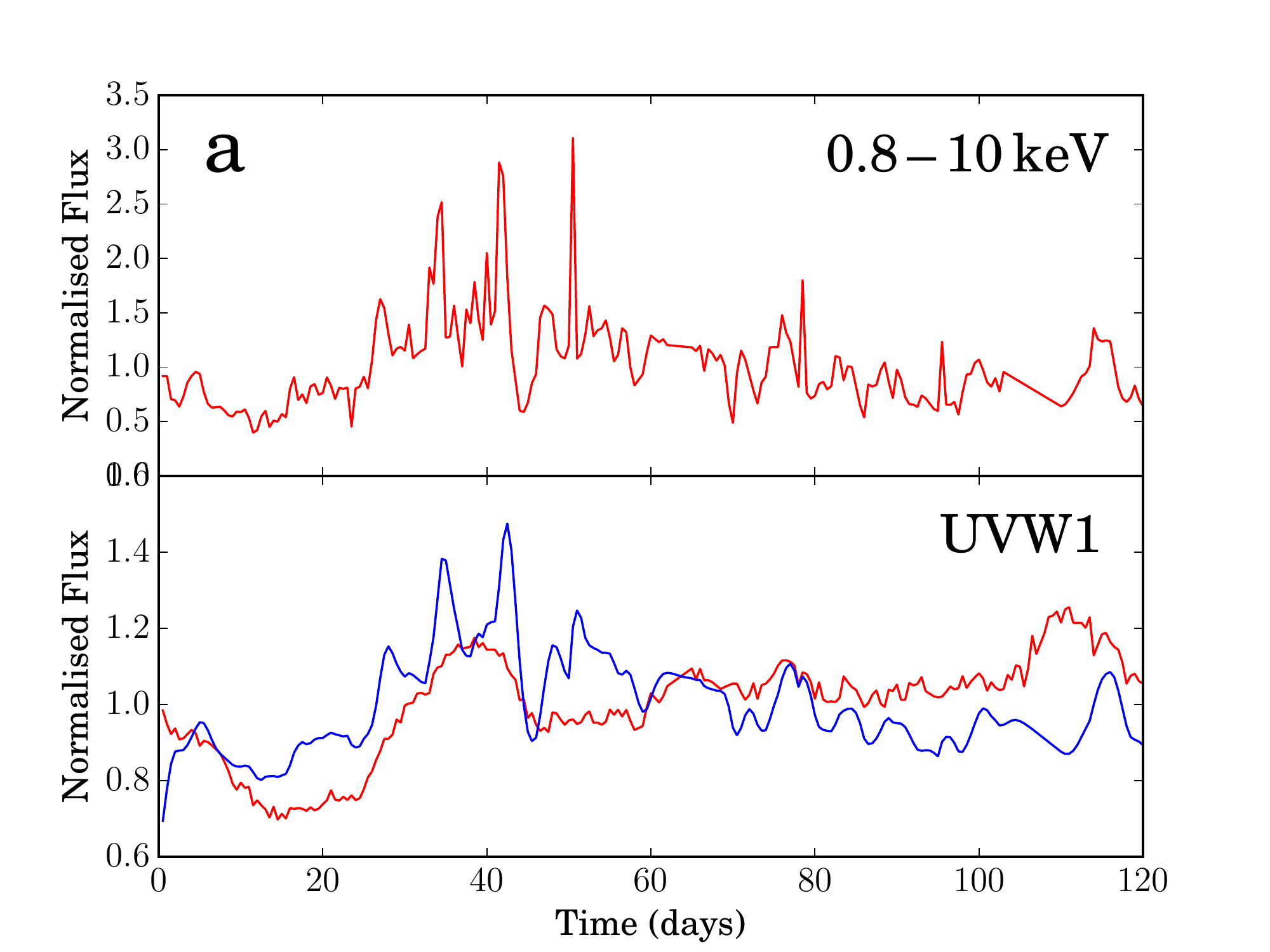} &
\includegraphics[width=8cm]{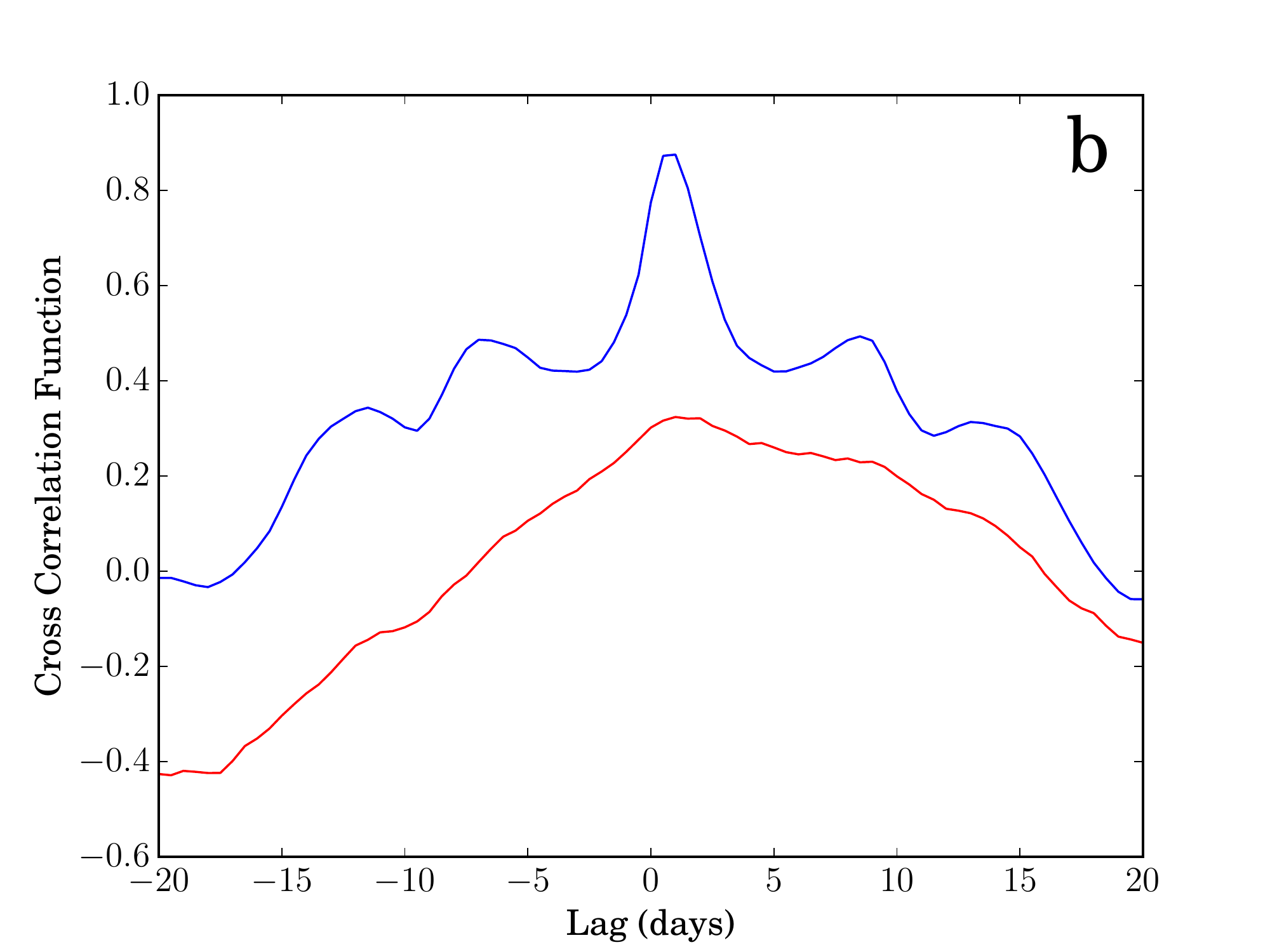} \\
\includegraphics[width=8cm]{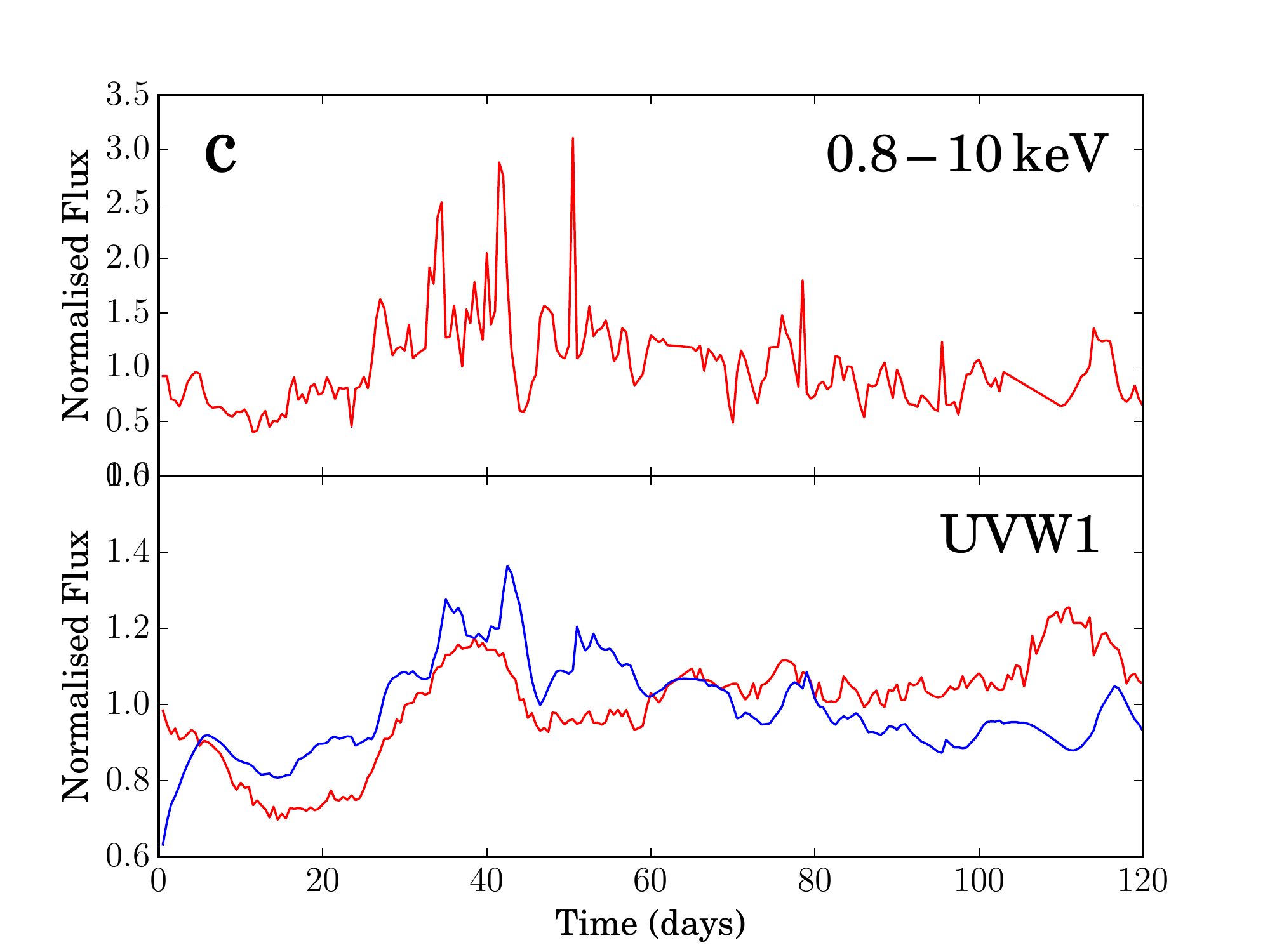} &
\includegraphics[width=8cm]{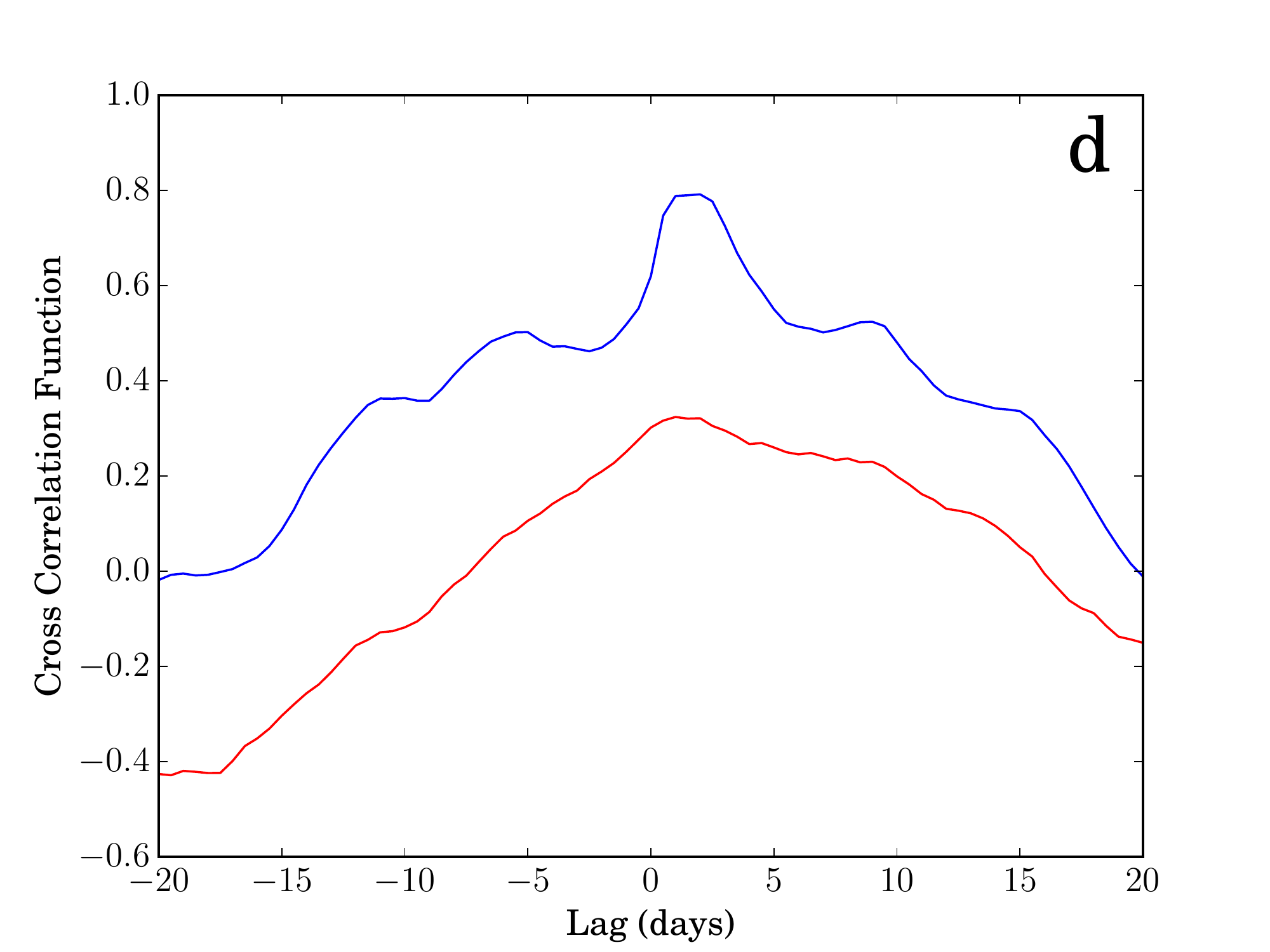} \\
\end{tabular}
\caption{BB disc with $r_{cor}=200$ reprocessing hard X-ray emission but with reprocessing coming from a larger effective radius. (a)\&(b) Comparison of simulated UVW1 lightcurve and CCF (blue) with the observations (red) for an effective reprocessing radius twice that expected from a standard BB disc. (c)\&(d) Comparison of simulated UVW1 lightcurve and CCF (blue) with the observations (red) for an effective reprocessing radius four times that expected from a standard BB disc.}
\label{fig2.1}
\end{figure*}

For our coronal power law fluctuations we use the hard X-ray light
curve of NGC 5548 presented by Edelson et al. (2015), interpolated
as described in the previous subsection to produce an evenly sampled
input lightcurve with $dt=0.5$\,d. We input this into our disc
reprocessing model and this allows us to calculate a model UVW1 light
curve, which we can then compare to the observed data. We show our
results in Fig.\ref{fig1}c (bottom panel), where the red line shows
the observed UVW1 lightcurve and the blue line shows our predicted
UVW1 lightcurve using this model. For reference we also show the
input hard X-ray lightcurve (top panel). Our simulated UVW1 light
curve clearly fails to reproduce both the amplitude 
of fast variability (much more in the model than in the data) and the overall
long term shape of the observed UVW1 lightcurve (especially the dip in the 
observed lightcurve at 18 days and the rise at 110 days). 

In Fig.\ref{fig1}d we show the cross correlation function (CCF) of the
simulated UVW1 lightcurve with respect to the hard X-ray lightcurve
(blue), compared to the CCF of the observed UVW1 lightcurve with
respect to the hard X-ray lightcurve (red). A positive lag indicates
the UVW1 band lagging the hard X-rays. The predicted CCF is strongly
peaked with almost perfect correlation at close to zero lag, and is
quite symmetric. The observed CCF has none of the well correlated,
narrow component at lags $<1$~day, but is instead quite poorly
correlated, and asymmetric with a peak indicating that the UVW1 band
lags behind the hard X-rays by $\sim0.5-2$\,d (Edelson et al. 2015).

Clearly, this is not a viable model of the UVW1 lightcurve. 
The data require more reprocessing at longer lags, and less
reprocessing at shorter lags. Certainly the inner disc produces the
shortest time lags, so truncating the BB disc at a larger radius could supress
some of the fast variability. We explore this in the next section.

\subsection{Increasing Disc Truncation}

We rerun our model with the BB disc truncated at a much larger radius,
such that $r_{cor}=200$. Such a large truncation radius for such
a low mass accretion rate AGN is consistent with the observed trend in
local AGN for $r_{cor}$ to anticorrelate with $L/L_{Edd}$ (Jin et al
2012; Done et al 2012). We show the results of using this larger truncation radius in
Fig.\ref{fig2}. The key difference in the spectrum 
is that the hottest parts of the BB disc
are no longer present, with the energy instead giving a slight increase in the
normalisation of the hard X-ray Comptonisation component. The small disc
component gives a significantly
lower UVW1 flux, so the variable components from both the direct
Comptonisation component and its reprocessed UV flux now contribute a higher fraction of 
the UVW1 band.  

The model can now better reproduce
the observed amplitude of UVW1 band long term flux variations (blue lightcurve
in Fig.\ref{fig2}c, especially the dip at 18 days). 
However our simulated UVW1 lightcurve still has much
more fast variability than is seen in the real UVW1 lightcurve. The
simulated UVW1 lightcurve clearly looks like the hard X-ray light
curve from which it was produced. Yet the observed UVW1 lightcurve
looks quite different to the observed hard X-ray lightcurve.

This is shown clearly in Fig.\ref{fig2}d where we compare the
CCFs. The CCF peak from our simulated lightcurve has shifted to
$\sim0.5$\,d rather than the close to zero-peaked CCF of the previous
model, but the lag is still not as long as in the observed CCF, and
the model UVW1 lightcurve is much more correlated with the hard X-ray
lightcurve. This is seen at all lags, but the problem is especially
evident on short lag times, showing quantitatively that the model UVW1
lightcurve has much more of the fast variability seen in the hard X-ray
lightcurve than the real data.

Edelson et al. (2015) commented that the lags they measure are much
longer than expected from reprocessing on a standard BB disc, i.e. the
radii that should show peak emission in the UVW1 band are much smaller
than the radii implied by the light travel time delayed response of
emission in that band. This would suggest that the accretion disc
around NGC 5548 is not a standard BB disc. Somehow the same emission is
produced at a larger radius than standard BB disc models predict.

We test this by altering our disc transfer functions such that the
reprocessing effectively occurs at twice (Fig.\ref{fig2.1}a\&b) and
then four times (Fig.\ref{fig2.1}c\&d) the radius at which a standard
BB disc would produce that emission. In theory this should improve our
simulated lightcurves, as reprocessing at larger radii smooths out
fast fluctuations so should reduce the amount of high frequency power
in the lightcurve.

\begin{figure*} 
\centering
\begin{tabular}{l|r}
\leavevmode  
\includegraphics[width=9cm]{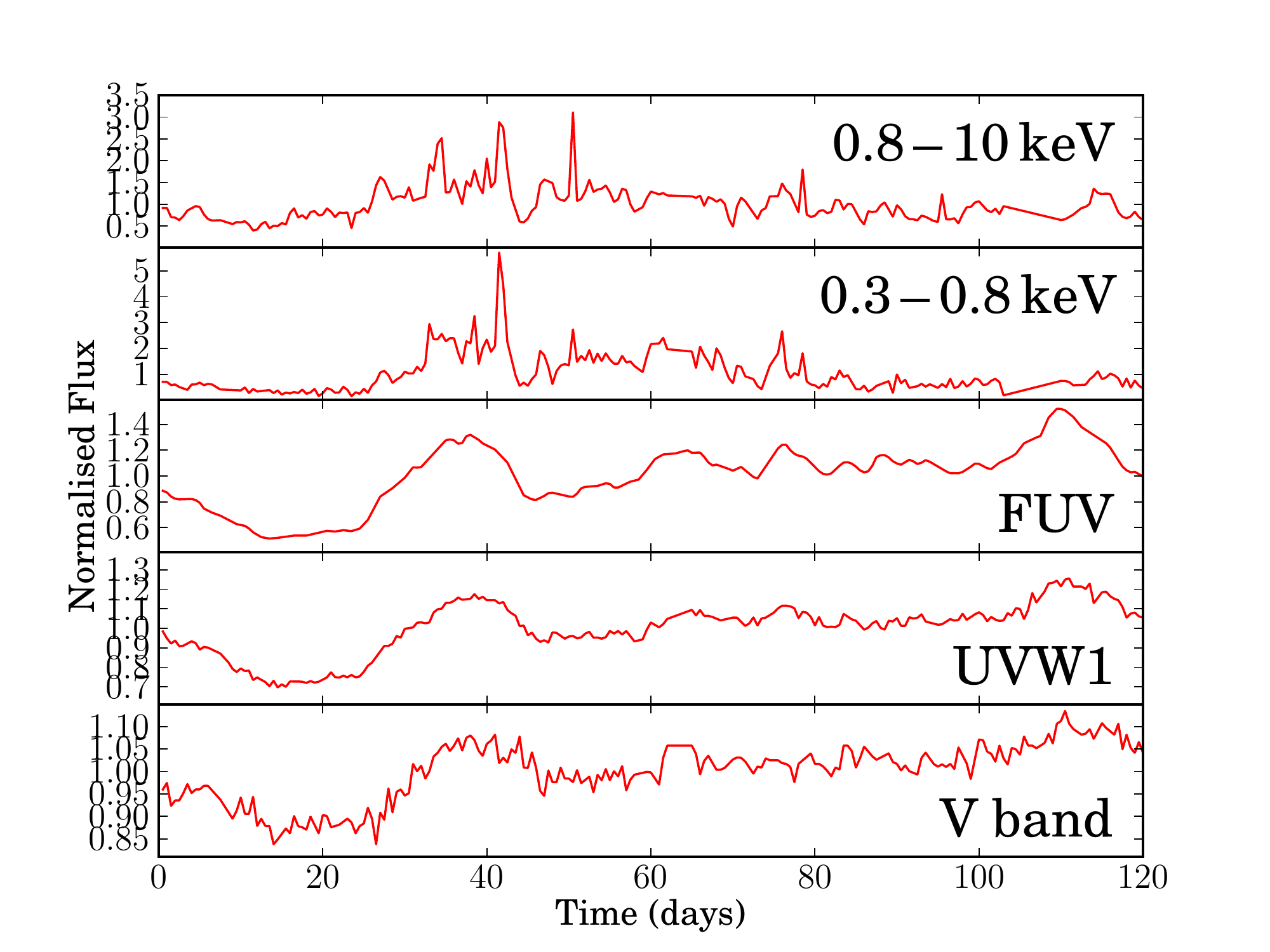} &
\hspace{-0.47in}
\includegraphics[width=9cm]{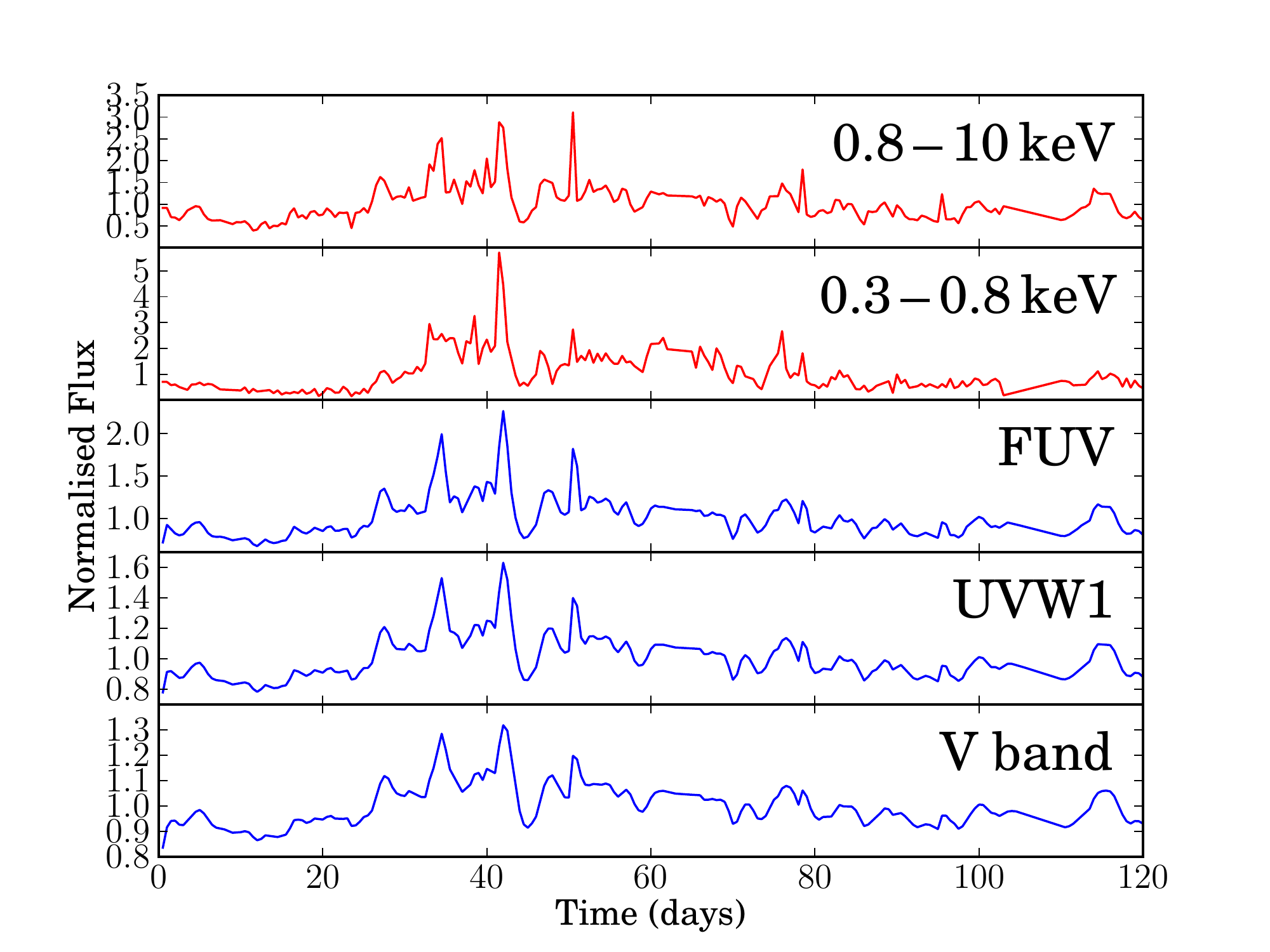} \\
\end{tabular}
\caption{Left panel: observed simultaneous hard X-ray, soft X-ray, FUV, UVW1 and V band lightcurves of NGC 5548. Right panel: observed simultaneous hard X-ray (input into the model) and soft X-ray lightcurves of NGC 5548, together with resulting FUV, UVW1 and V band lightcurves simulated by model with a standard BB disc with $r_{cor}=200$ reprocessing hard X-ray emission.}
\label{fig3}
\end{figure*}

However, Fig.\ref{fig2.1}a shows that an effective radius twice that
of a standard BB disc does not smooth the simulated lightcurve enough. An
effective radius four times that of a standard BB disc does a better job
(Fig.\ref{fig2.1}c), but comparison of the CCFs (Fig.\ref{fig2.1}d)
illustrates that this is still not a good match to the data. An
effective radius four times that of a standard BB disc may be required to
sufficiently reduce the high frequency power in the lightcurve, but
this then gives light travel time lags that are too long. The peak lag
of the observed CCF is roughly $0.5-2$\,d, while the peak lag from the
model lightcurve extends from $\sim1-3$\,d. 
The observed smoothing timescale is much longer than the lag
timescale, and this cannot be replicated by light travel time
smoothing, as light travel time effectively ties the smoothing to the 
lag timescale. For this reason, doubling the black hole mass to match the
Bentz et al. (2010) estimate does not solve the problem of
transmitting too much high frequency power into the optical. Increasing
the disc inclination angle increases the amount of smoothing, however this
effect is negligible even when setting $i=75^\circ$ (an unreasonably
large angle for a Seyfert 1 such as NGC 5548) and likewise cannot
reduce the high frequency power in the model optical lightcurves.

Moreover, the model lightcurves are all far more correlated
with the hard X-ray lightcurve, on all timescales, than the observed
lightcurve is. Even in Fig.\ref{fig2.1}d the peak correlation
coefficient between the hard X-rays and the model UVW1 lightcurve is
$\sim0.8$, while between the hard X-rays and the observed lightcurve
it is only $\sim0.3$.

Fig.\ref{fig3} further illustrates this. In the left panel we show the
observed lightcurves --- from top to bottom, hard X-rays, soft
X-rays, FUV, UVW1 and V band. In the right panel we again show the
observed hard and soft X-ray lightcurves, followed by our model
FUV, UVW1 and V band lightcurves using the standard BB disc model. It
is clear that reprocessing the hard X-ray lightcurve off a standard BB
disc size-limited by the self-gravity radius produces UV and optical
lightcurves that look like the original hard X-ray lightcurve. The
hard X-ray flares (e.g. at 42 days) are slightly more smoothed in the
longer wavelength bands, but they are still clearly recognisable.

By contrast the observed UV and optical lightcurves lack any short
term flares and show additional longer timescale variability that is
not present in the hard X-ray lightcurve (e.g. the dip between
$\sim10-30$\,d). The FUV to V band lightcurves are clearly well
correlated with one another, with peak CCFs with respect to the FUV of
$0.7-0.9$ (Edelson et al. 2015) and the increasing lags with
increasing wavelength suggest reprocessing is a key linking
factor. But they are much less well correlated with the hard X-rays
(with a peak CCF of only $\sim0.3$). The left panel of Fig.\ref{fig3}
shows there is a clear break in properties between the observed X-ray
and UV--optical lightcurves. The right panel shows that, if the hard
X-rays are the source of illuminating flux, this cannot occur. 

There are really only two ways out of this impasse. Either the hard
X-ray lightcurve is not a good tracer of the illuminating flux, or the
BB accretion disc is shielded from being illuminated by the hard X-rays. 
The former is plainly a possibility, as the hard X-ray lightcurve from
0.8-10~keV is not at the peak of the hard X-ray emission and Noda et al. (2011) show that there can be a fast variable steep power law component. Mehdipour et
al. (2015) and Ursini et al. (2015) show that the hard X-ray variability encompasses both a change in
normalisation {\it and} in spectral index, with the spectrum softening
as the source brightens. A hard X-ray lightcurve at 100~keV would be a
better direct tracer of the total hard X-ray flux, but until this is
available, we estimate this using the intrinsic power law spectral index
and normalisation derived by Mehdipour et
al. (2015). These are binned on 10 day intervals to get enough signal to
noise, so it is not a sensitive test of the model, but assuming that
the spectral cut-off remains fixed at 100~keV this still does not give a good
match to the FUV lightcurve. We conclude that it is more likely that
the reprocessing region is shielded from the hard X-ray illumination,
and consider below how this might also shed light on the origin of the
soft X-ray excess.

\section{Lightcurves and Spectra from Blackbody Disc Reprocessing FUV Emission}

\begin{figure*} 
\centering
\begin{tabular}{l|r}
\leavevmode
\includegraphics[width=8cm]{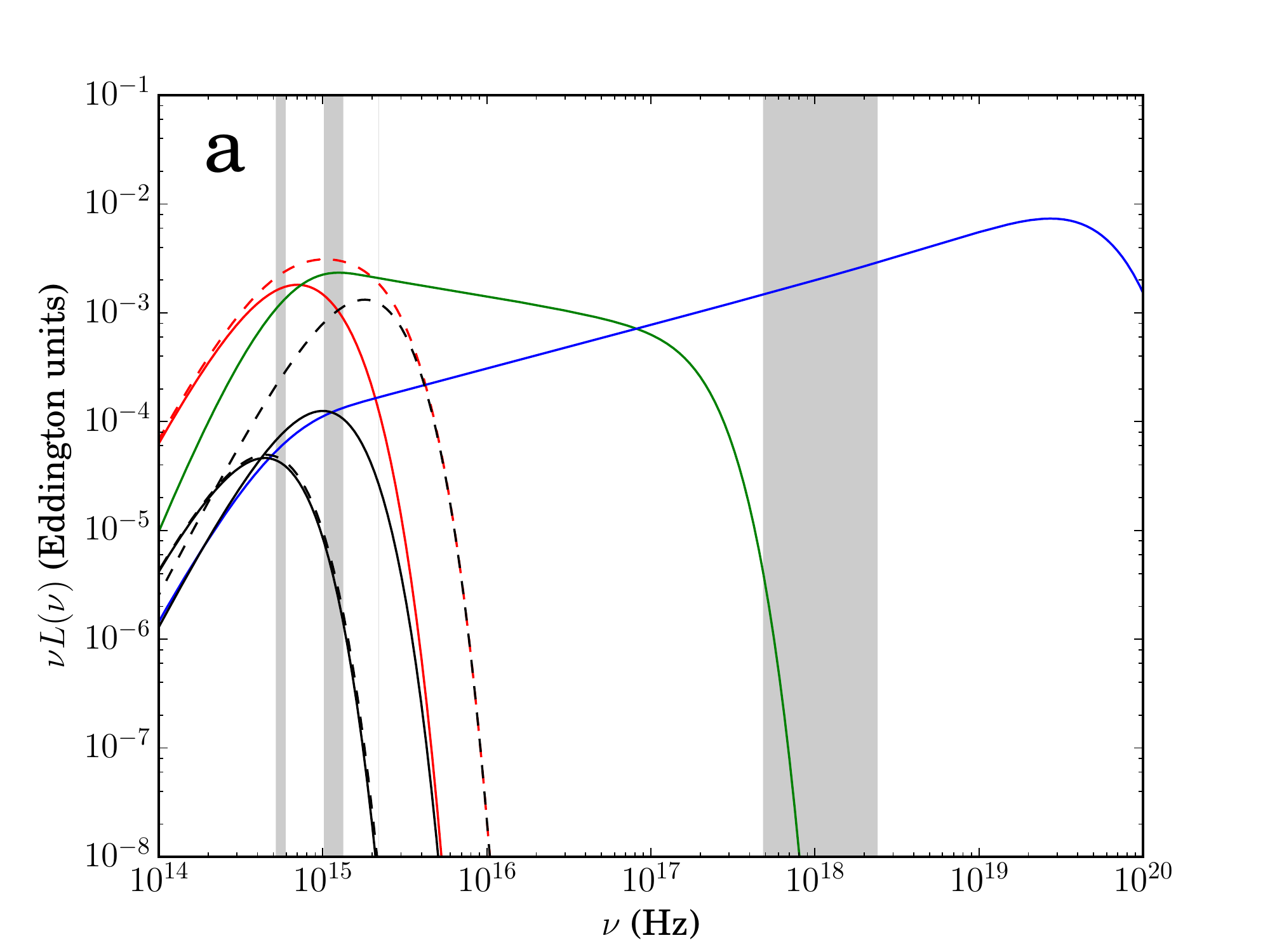} &
\includegraphics[width=8cm]{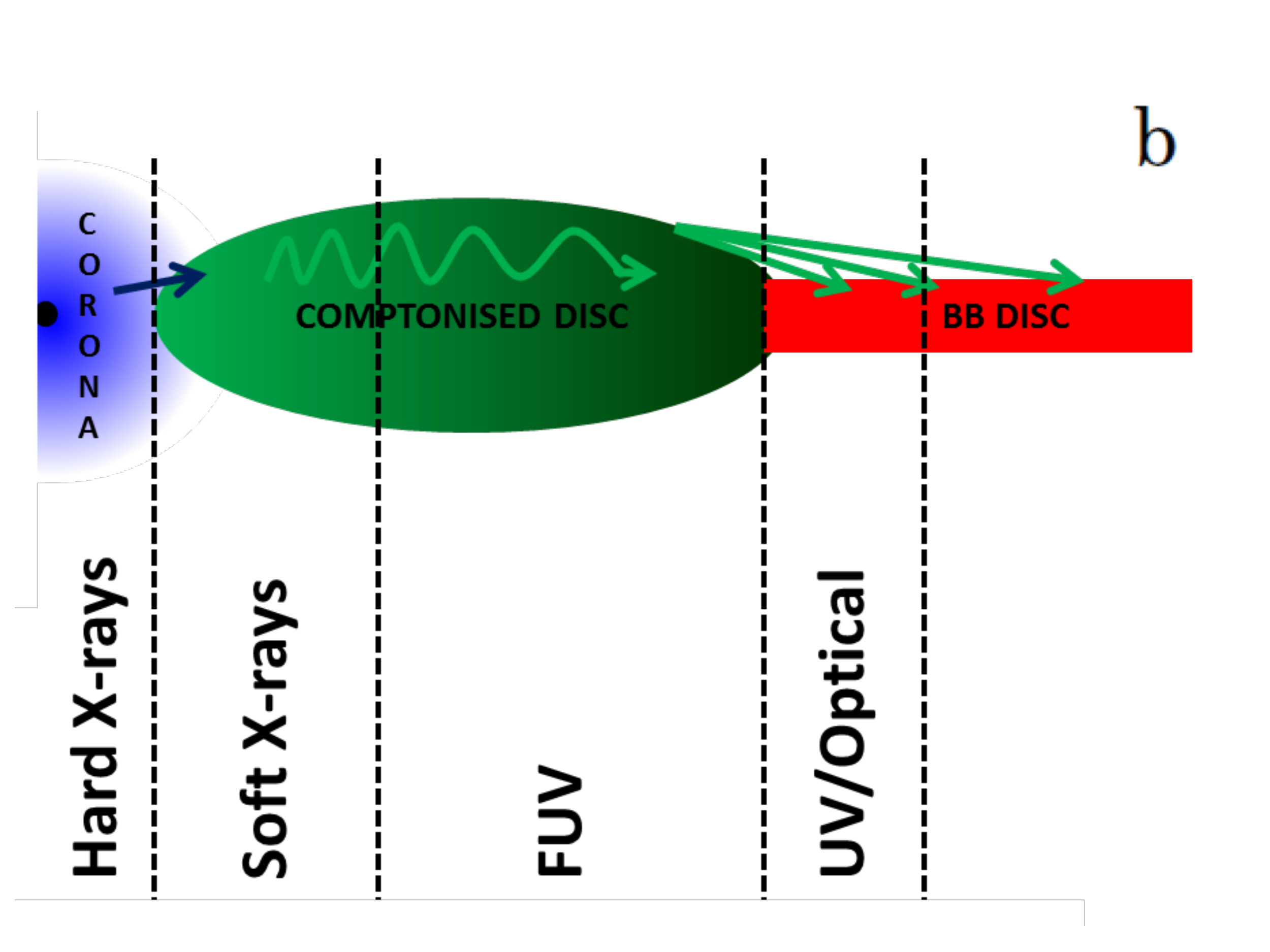} \\
\includegraphics[width=8cm]{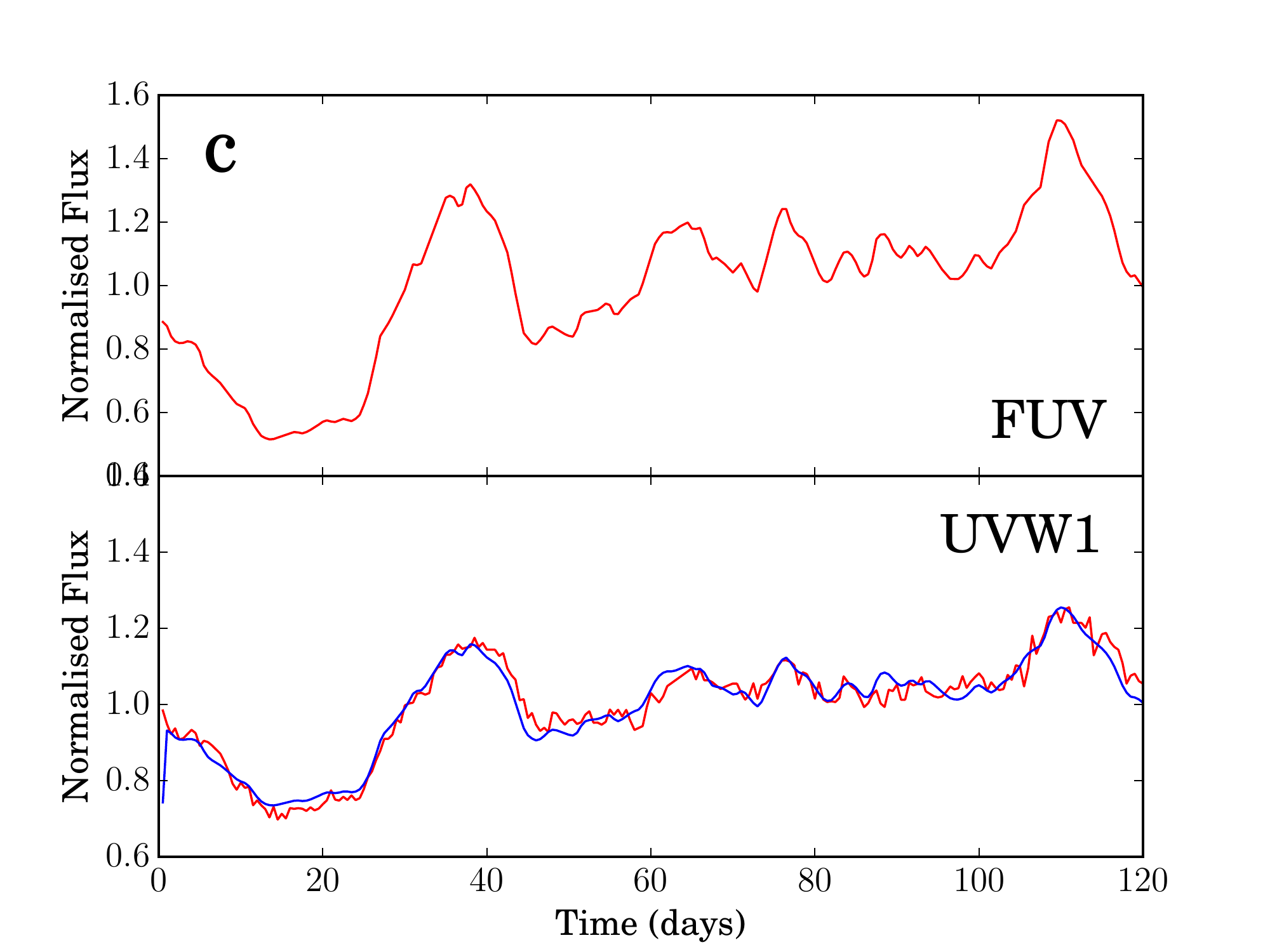} &
\includegraphics[width=8cm]{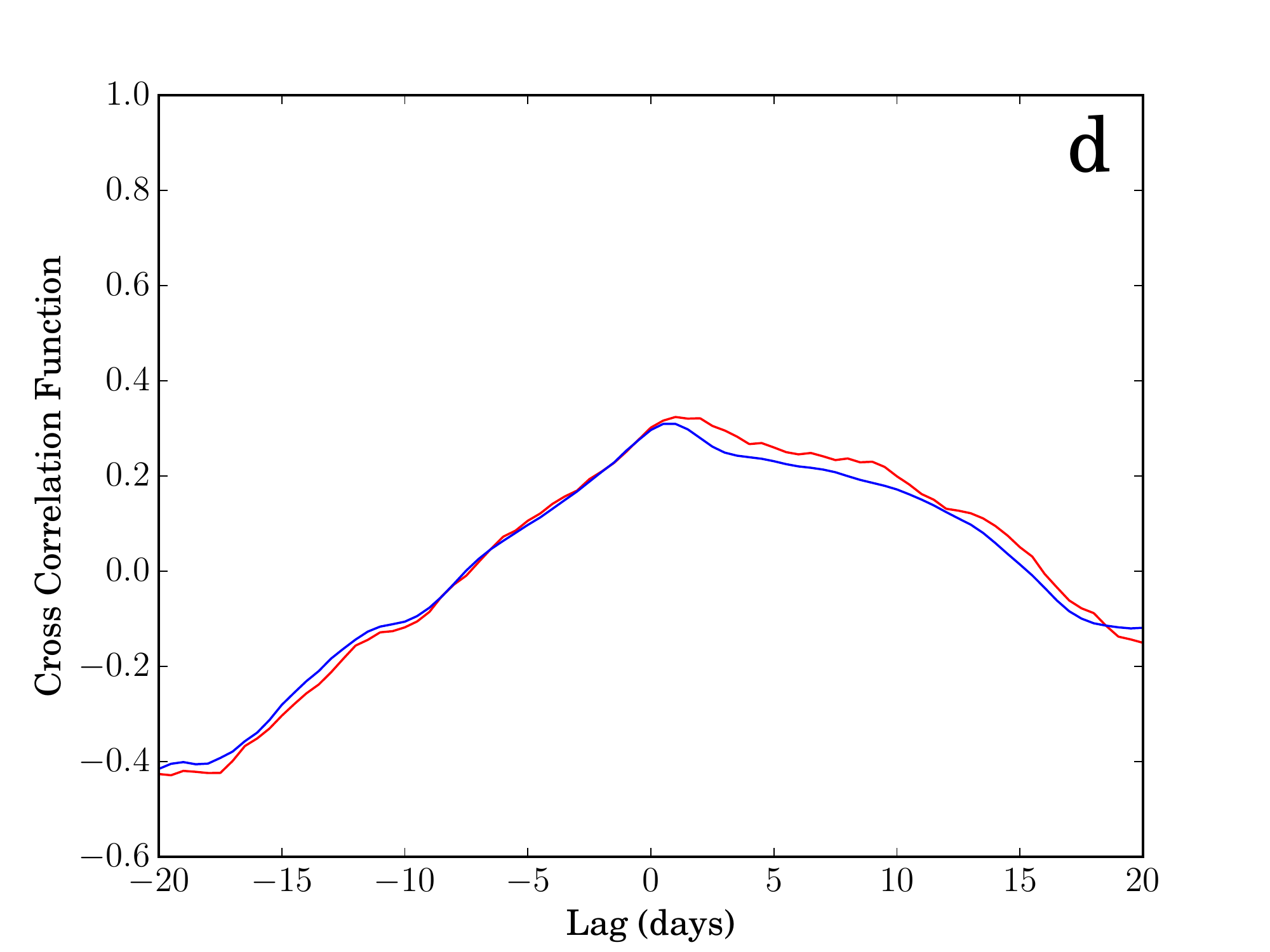} \\
\includegraphics[width=8cm]{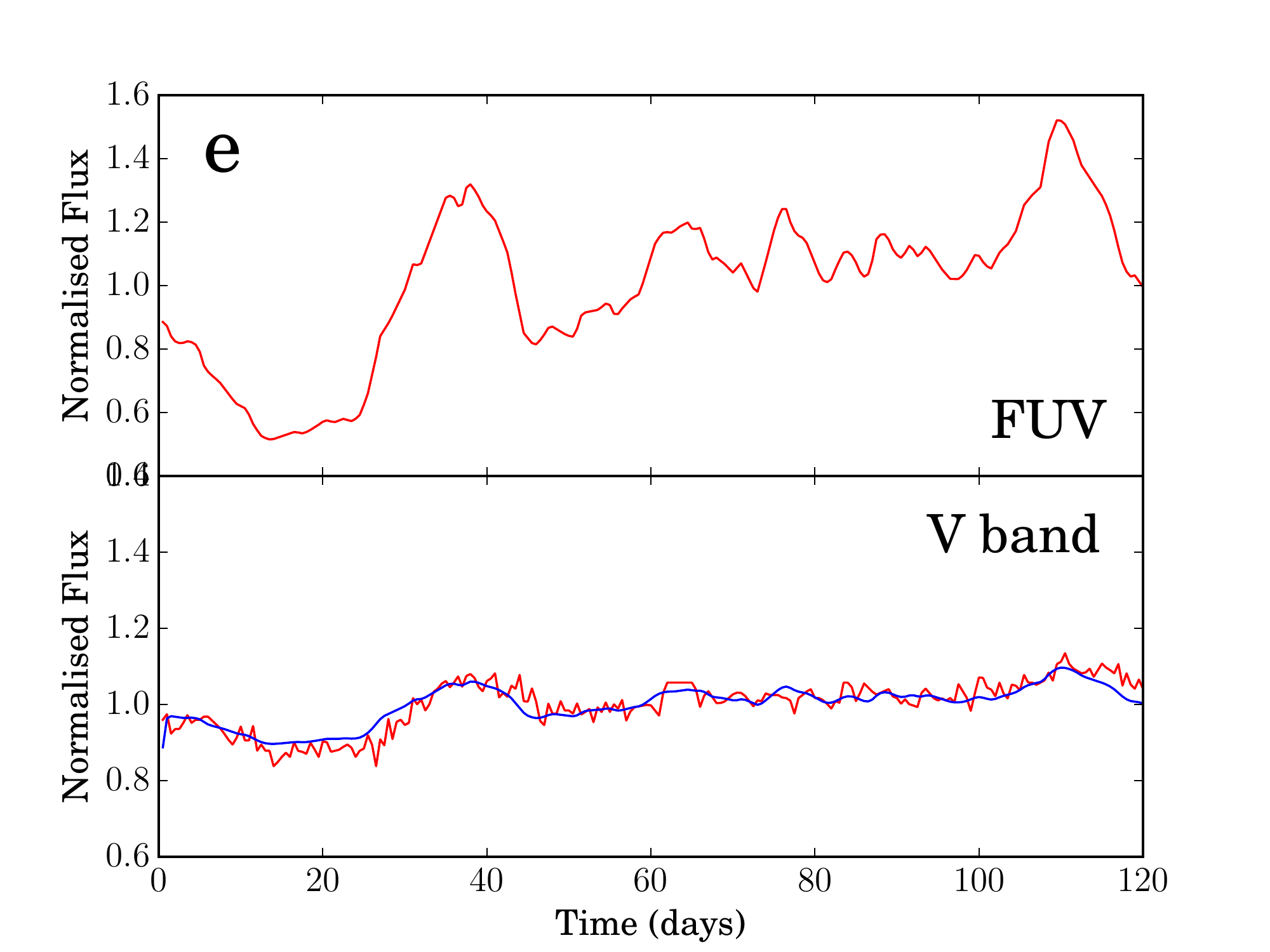} &
\includegraphics[width=8cm]{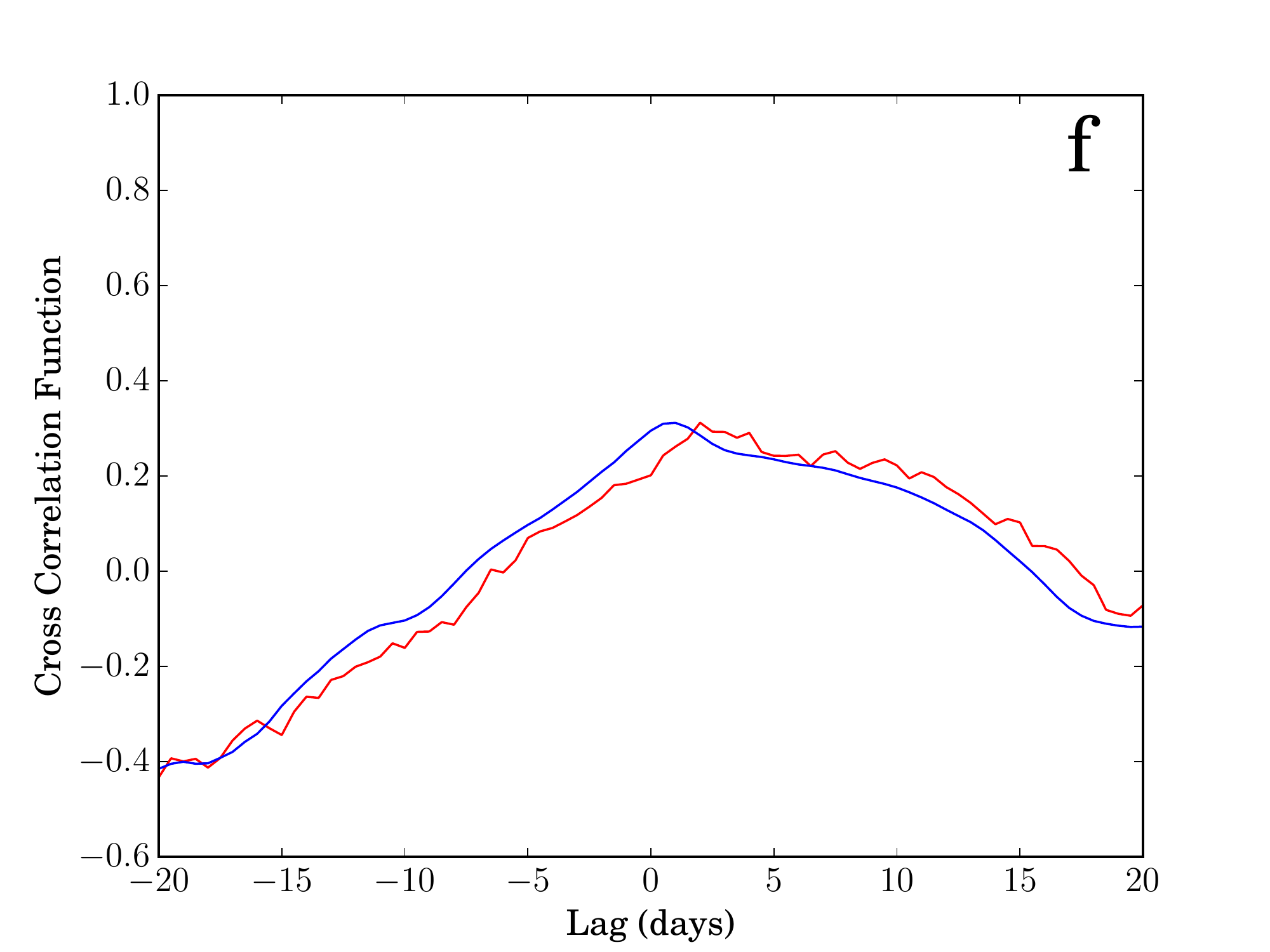} \\
\end{tabular}
\caption{Standard BB disc plus Comptonised disc component, with $r_{cor}=200$ and BB disc reprocessing FUV emission. (a) Model spectrum: solid red line shows total intrinsic BB disc emission; dashed red line shows total BB disc emission including reprocessing; black lines show emission from outermost and innermost BB disc radii, with solid and dashed lines for intrinsic and intrinsic plus reprocessed emission, respectively; green line shows optically thick Compton component from Comptonised disc; blue line shows hard coronal power law; grey shaded regions, from left to right, show location of V band, UVW1 band, FUV band and hard X-ray band respectively. (b) Cartoon of a scenario where the Comptonised disc obscures the inner hard X-ray region and is instead the source of flux illuminating the outer BB disc. Dashed lines suggest transitions between regions of material producing the bulk of the emission in a given band. (c) Top panel shows observed FUV lightcurve of NGC 5548 input into the model; bottom panel shows observed simultaneous UVW1 lightcurve (red) compared with simulated UVW1 lightcurve (blue). (d) Cross correlation function of observed UVW1 lightcurve with respect to observed hard X-ray lightcurve (red), and simulated UVW1 lightcurve with respect to observed hard X-ray lightcurve (blue). Positive lag values indicate the UVW1 band lagging behind the hard X-rays. (e)\&(f) same as (c)\&(d) but for observed and simulated V band lightcurves.}
\label{fig4}
\end{figure*}

The broadband spectrum of NGC 5548 presented by Mehdipour et al. (2015)
shows that a two component BB disc + hard power law model is clearly not
sufficient to fit its spectrum. This source shows a strong soft X-ray
excess above the $2-10$\,keV power law which can be well fit with an
additional low temperature, optically thick Compton component, though
this is not a unique interpretation. It can also be well fit in the
0.3-10~keV bandpass with highly smeared, partially ionized reflection
(Crummy et al. 2006). However, with the advent of NuSTAR and other
high energy instruments, it is now clear that the reflection
interpretation does not give such a good fit to the data up to
50-100~keV for AGN with hard X-ray spectra (e.g. Matt et al. 2014; 
Boissay et al. 2014). Hence we assume that the optically thick Compton component, 
which produces the soft X-ray excess, is an
additional intrinsic continuum component. 

To include this, we now assume that the BB disc truncates at $r_{cor}$ as
before, but that the remaining gravitational power inwards of this
radius is split between two coronal components: the hard power law and
the optically thick Compton component. We fix the temperature and
optical depth of the optically thick Compton component to $kT_e =
0.17$\,keV and $\tau=21$ (Mehdipour et al. 2015). This component then
peaks in the UV as required by the spectrum. By reducing the fraction
of coronal energy in the hard power law to $f_{pl}=0.75$, i.e. $0.25$ of
the coronal energy goes instead into powering the optically thick Compton
component, we are again able to match the ratio of
$F_{UVW1}/F_{10{\rm{\,keV}}}\sim1.7$ found by Mehdipour et al. (2015),
with $r_{cor}=200$, which we could not do previously with a
substantially truncated disc and no optically thick Compton component. In
Fig.\ref{fig4}a we show this new model spectrum, with the optically thick Compton
component shown in green.

While the origin of this emission is not well understood, it is
clearly not from a standard disc. This component takes over from the
standard BB disc in the UV, which may not be a coincidence as 
the substantial atomic opacity in the UV can cause changes in the disc
structure compared to a Shakura-Sunyaev disc which incorporates only plasma
opacities of electron scattering and free-free absorption.
In particular, UV line driving has the potential to lift the disc photosphere
(e.g. Laor \& Davis 2014). The copious hard X-ray emission in
this object should quickly over-ionise any potential UV line driven
wind, which would result in the material falling back down again
without being expelled from the system. This scenario has the
potential to effectively increase the scale-height of the disc,
decreasing its density (Jiang, Davis \& Stone 2016). This decreases its true opacity, hence
increasing the effective colour temperature correction (e.g. Done et
al. 2012). Alternatively, UV temperatures are also linked to the onset of the 
dramatic disc instability connected to hydrogen ionisation (Lasota 2001; Hameury et al. 2009; 
Coleman et al 2016). Whatever the origin, the 
total optical depth of the disc to electron scattering at the UV radii is
probably of order 10-100 (Laor \& Netzer 1989), so the optically thick Compton
emission in this picture is coming from the disc itself, with the emission not
quite able to thermalise to standard BB emission because of the increased scale-height of the
disc due to the UV radiation pressure and/or the onset of the
hydrogen ionisation instability. We therefore refer to the optically thick Compton component as the `Comptonised disc', to distinguish it from the outer BB disc.

\begin{figure} 
\centering
\begin{tabular}{l}
\leavevmode
\includegraphics[width=8cm]{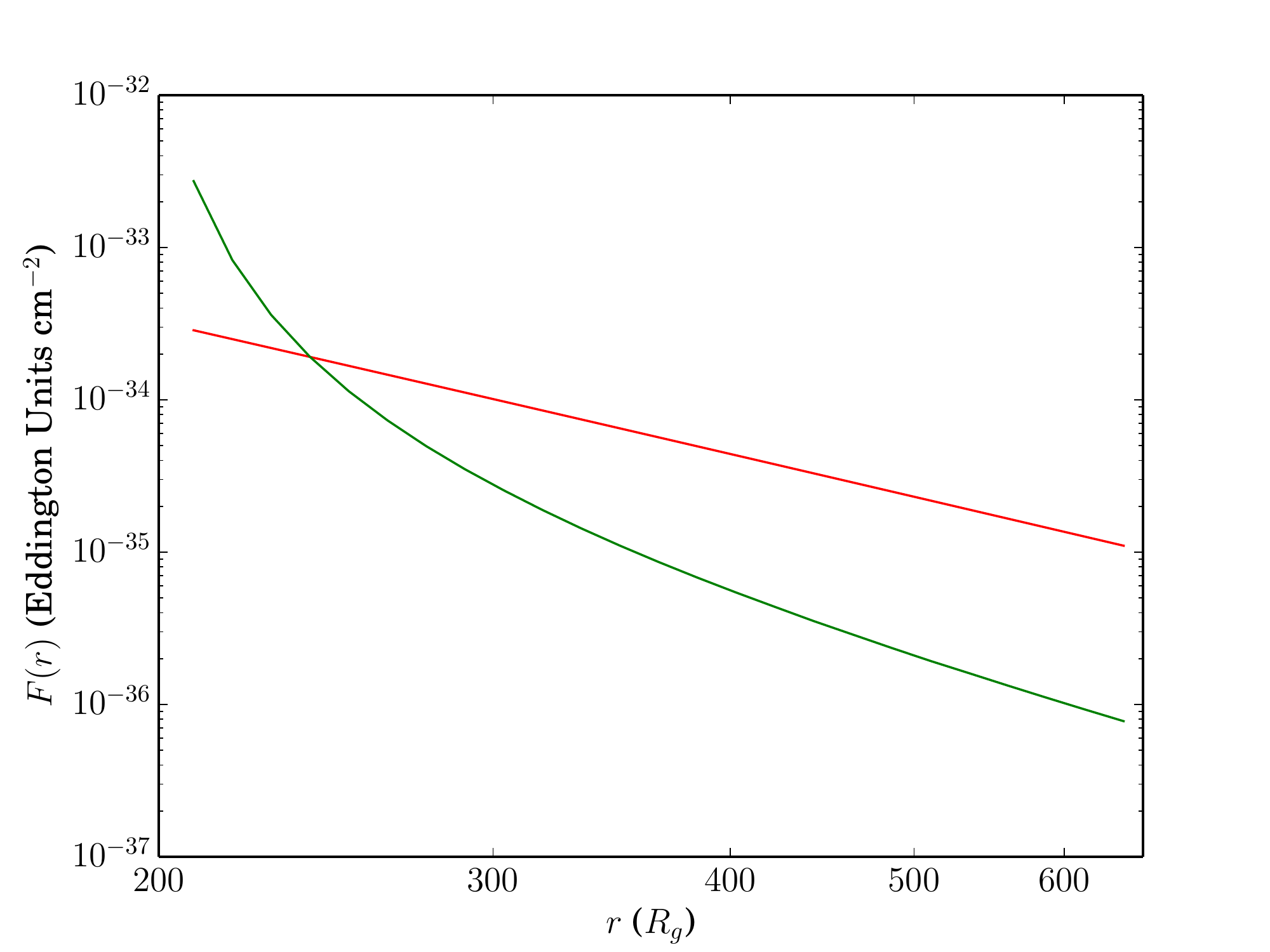} \\
\end{tabular}
\caption{Illuminating FUV flux as a function of BB disc radius (for a flat disc) assuming the FUV flux is emitted by a wall of material of height $10\,R_g$ located at $200\,R_g$ (green line). Red line shows gravitational flux dissipation of BB disc.}
\label{fig4_0}
\end{figure}

The break in properties between the soft X-ray and FUV
lightcurve show clearly that the Comptonised disc is itself
stratified rather than being a single spectral component as in
Mehdipour et al. (2015). We show our potential geometry in
Fig.\ref{fig4}b, where the soft X-ray emission comes from the inner
regions of this large scale-height flow, which still cannot illuminate
the outer BB disc, while the FUV is produced at larger radii and can
illuminate the outer BB disc. 

We model this FUV illumination by assuming a cylinder of material
located at a particular radius ($r_{irr}>0$), with a particular height
($h_{max}$). This changes the illumination pattern, so we can no
longer use the approximation of an on-axis point source. We calculate
the reprocessed flux at a given BB disc radius by dividing the `surface'
of the wall into elements (azimuthally --- $d\phi$ --- and vertically
--- $dh$) and summing the flux contribution from each element:

\begin{equation}
F_{rep}(r) = \int_0^{h_{max}}\int_0^{2\pi} \frac{f_{irr}L_{cor}}{2\pi^2h_{max}R_g^2} \frac{h \,dh \,d\phi}{(r^2+r_{irr}^2+h^2-2r_{irr}r\cos\phi)^{3/2}}
\end{equation}

This new illumination pattern intensifies the illumination on radii
close to $r_{irr}$ but for $r\gg r_{irr}$ it becomes indistinguishable
from the case of a lamppost point source. We set $r_{irr}=r_{cor}$ and
$h_{max}=10$, implying a scale-height for the outer edge of the
Comptonised disc of $h/r=0.05$, which should be sufficient to obscure the
hard X-ray emission. The resulting illumination profile is shown in
Fig.\ref{fig4_0} (green line).

In Fig.\ref{fig4}c\&d we compare the resulting model UVW1 lightcurve
with the observations. We now find a much better match to the
behaviour of the observed UVW1 lightcurve, matching the amplitude of
fluctuations and reproducing the shape of the UVW1/hard X-ray CCF. 

\begin{figure} 
\centering
\begin{tabular}{l}
\leavevmode
\includegraphics[width=8cm]{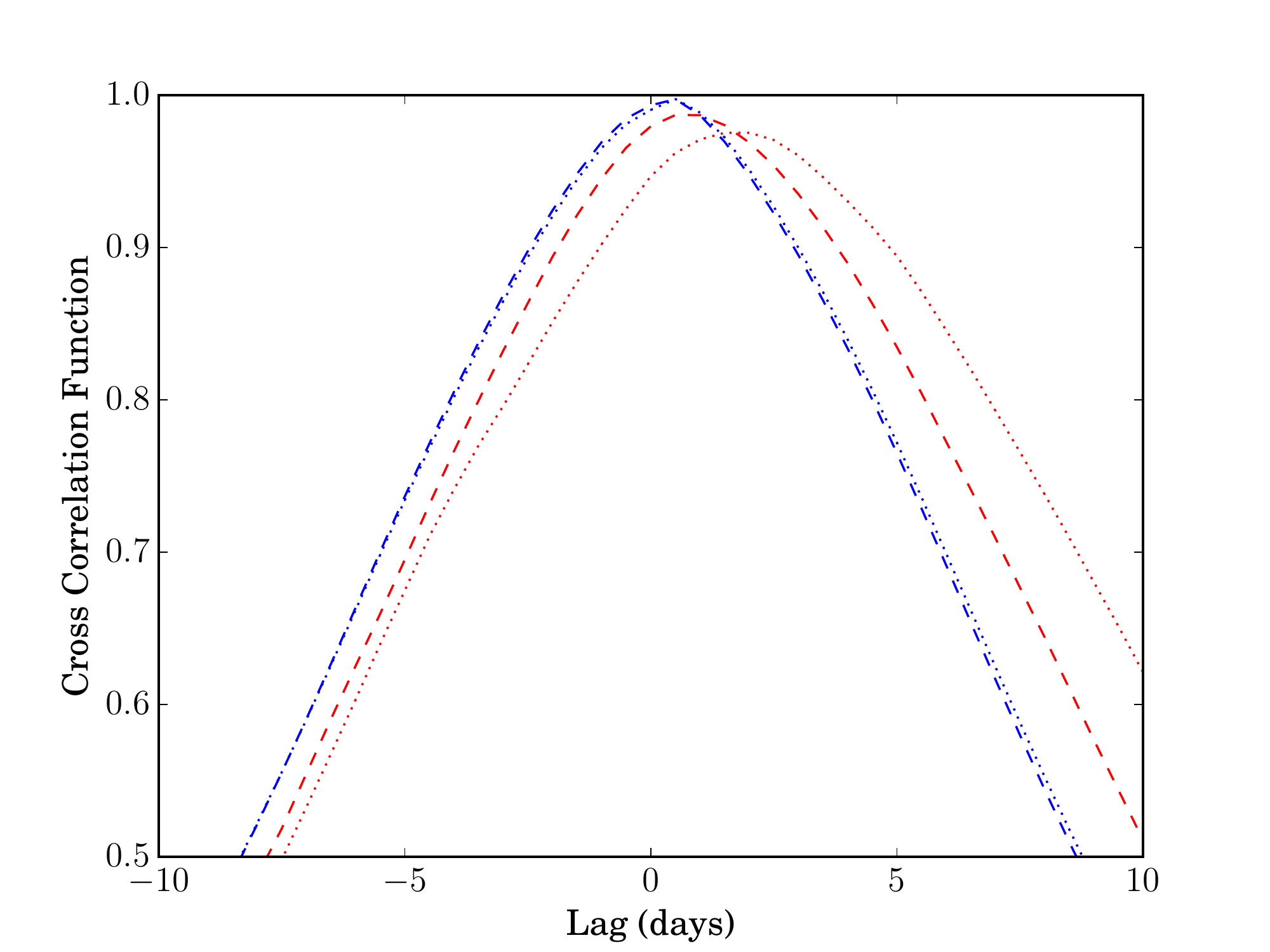} \\
\end{tabular}
\caption{Comparison of cross correlation functions with respect to the FUV for the standard BB disc plus Comptonised disc FUV reprocessing model. Dashed lines show CCF of UVW1 with respect to FUV, and dotted lines show CCF of V band with respect to FUV. Red lines show CCFs calculated using the observed lightcurves, while blue lines show the corresponding CCFs calculated using the simulated lightcurves.}
\label{fig5a}
\end{figure}

In Fig.\ref{fig4}e\&f we also compare our model V band lightcurve
with the data. We find a good match to the amplitude of the observed
fluctuations, but the model V band CCF with respect to the hard X-rays (blue) is not lagging by as much as the real
V band lightcurve (red). This mismatch is more evident when comparing
the model UVW1 and V band CCFs with respect to the FUV. Fig.\ref{fig5a} shows that
both the model UVW1 (blue dashed) and V band (blue dotted) CCFs
with respect to the FUV peak at close to zero lag, while the real UVW1 (red
dashed) and V band (red dotted) CCFs are significantly
shifted away from zero lag. 

Thus while reprocessing the FUV gives lightcurves which are a much
better match to the data, the response of our model lightcurves is too
fast. The observed V band lag behind the FUV lightcurve is $\sim2$\,d
(Edelson et al 2015). For a black hole mass of
$3.2\times10^7\,M_\odot$, this means the reprocessed V band flux must
be emitted roughly $1080\,R_g$ away from wherever the FUV emission
occurs. The observed UVW1 lag is $\sim0.5$\,d, implying it is
emitted at a distance of $\sim270\,R_g$ from the FUV emission. In our
model we assume the FUV emission is supplied by the outer edge of the
Comptonised disc at $\sim200\,R_g$. However, regardless of the exact
location of the FUV emission, the observed lightcurves imply that the
reprocessed V band flux must be emitted $\sim700\,R_g$ further away
from the FUV continuum than the reprocessed UVW1 flux. So far we have
used a BB disc truncated at the self-gravity radius of $660\,R_g$ as our
reprocessor. Clearly this does not provide a large enough span of
reprocessing radii. We now rerun our model with a larger outer radius 
for the BB disc, to see if we can reproduce the observed length of the V band
lag.

\begin{figure} 
\centering
\begin{tabular}{l}
\leavevmode
\includegraphics[width=8cm]{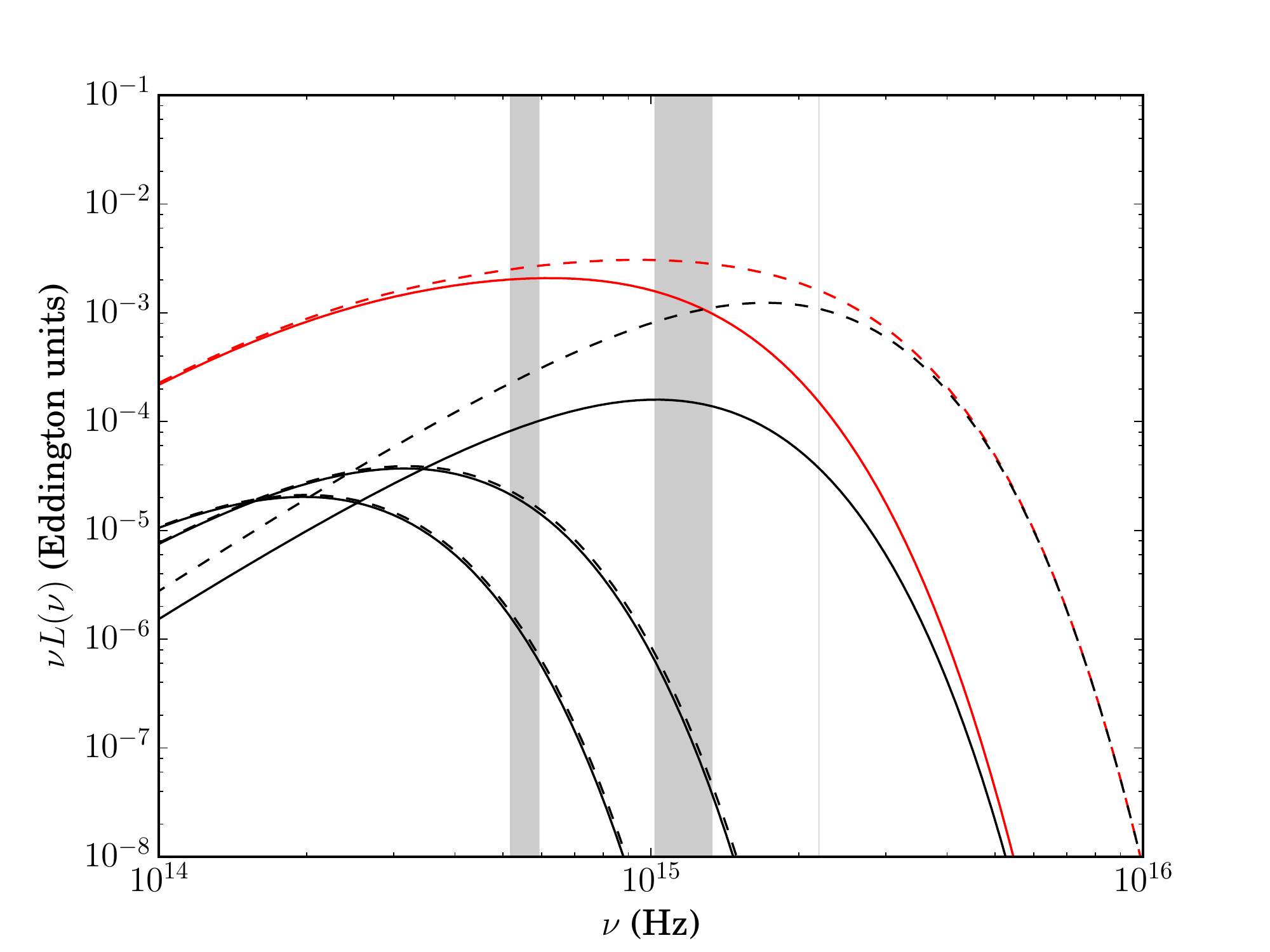} \\
\end{tabular}
\caption{Disc spectrum from the standard BB disc reprocessing FUV emission model with $r_{cor}=200$ and $r_{out}$ increased to $2000$. Solid red line shows total intrinsic BB disc emission; dashed red line shows total BB disc emission including reprocessing; pairs of black lines from left to right show emission from outermost radius, an intermediate radius ($r=1000$) and innermost radius of the BB disc, with solid and dashed lines for intrinsic and intrinsic plus reprocessed emission respectively. Grey shaded regions, from left to right, show location of V band, UVW1 band and FUV band, respectively.}
\label{fig5}
\end{figure}

\begin{figure*} 
\centering
\begin{tabular}{l|r}
\leavevmode
\includegraphics[width=8cm]{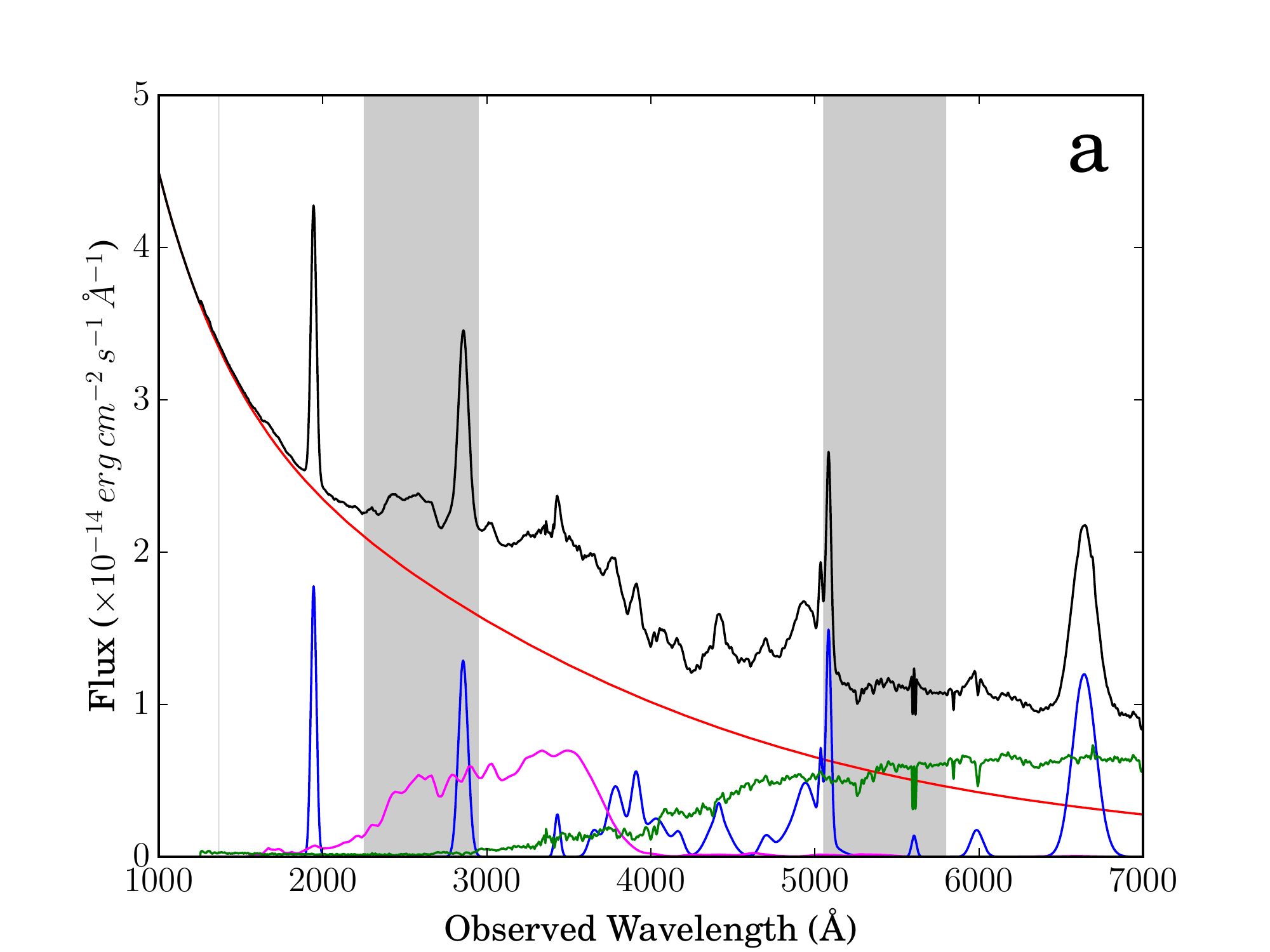} &
\includegraphics[width=8cm]{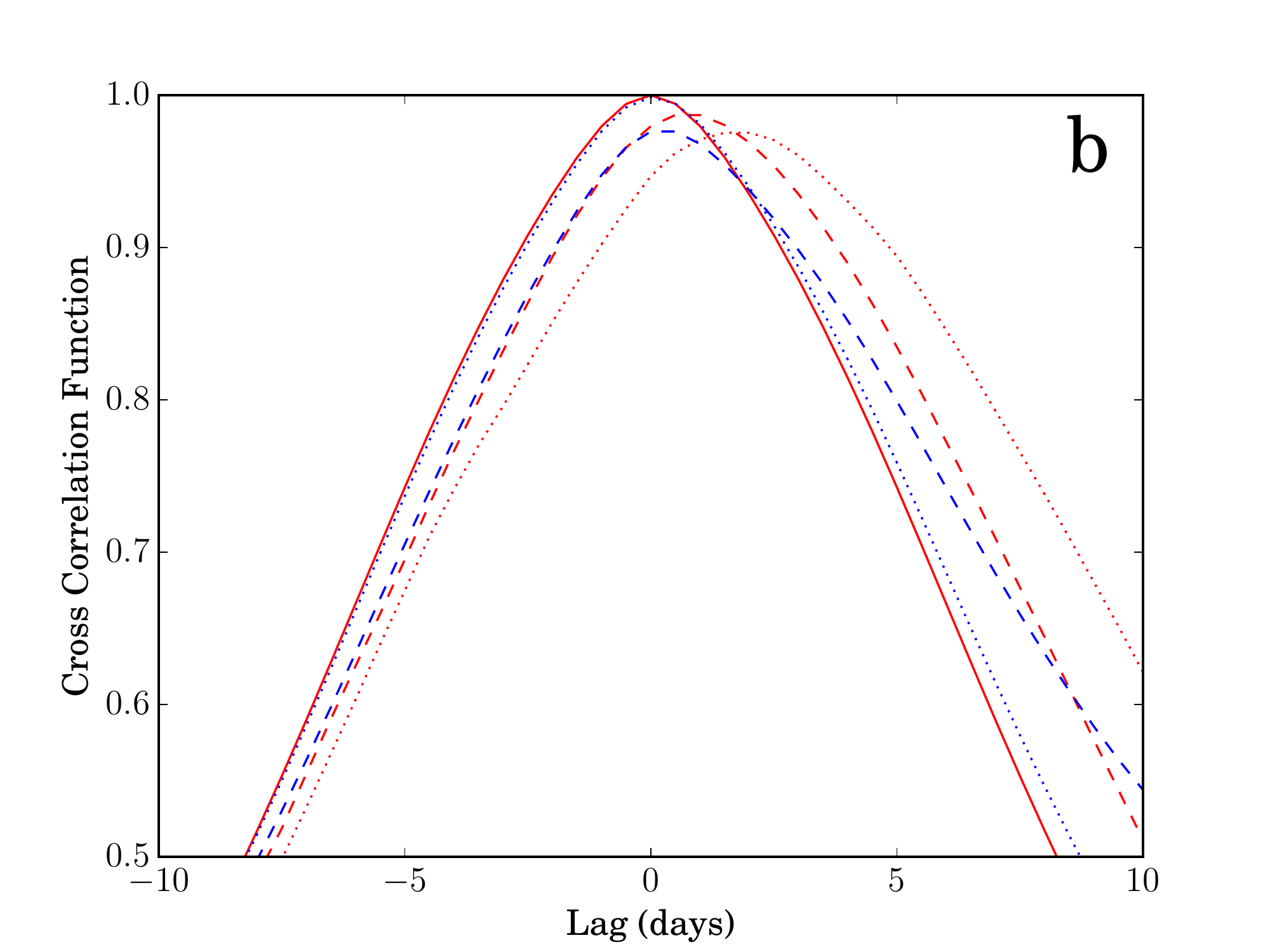} \\
\end{tabular}
\caption{Model assuming line emission is the only source of reprocessed emission. (a) UV/Optical spectral fit for NGC 5548 from Mehdipour et al. 2015, together with their spectral decomposition. Red line shows continuum emission, blue shows broad and narrow line components, magenta shows blended Fe{\sc{II}} with Balmer continuum, green shows host galaxy contribution and black shows resulting total spectrum. Grey shaded regions, from left to right, show location of FUV, UVW1 and V bands, respectively. (b) Comparison of cross correlation functions for model assuming line emission is the only source of reprocessed emission. Solid line shows CCF of FUV with respect to FUV (i.e. autocorrelation), dashed lines show CCF of UVW1 with respect to FUV, and dotted lines show CCF of V band with respect to FUV. Red lines show CCFs calculated using the observed lightcurves, while blue lines show the corresponding CCFs calculated using the simulated lightcurves.}
\label{fig6}
\end{figure*}

\subsection{Increasing Outer Disc Radius}

We increase the outer radius of the BB disc to $2000\,R_g$ and rerun our
FUV reprocessing simulation. We find that increasing the outer radius
of the BB disc makes no difference to the length of our simulated V band
lag. Fig.\ref{fig5} illustrates why.

A $2\,$d V band lag requires the V band to be dominated by reprocessed
flux emitted from $R=2\,ld>1000\,R_g$ (for a $3.2\times10^7\,M_\odot$ like
NGC 5548). Fig.\ref{fig5} shows that BB disc annuli at these large radii
are simply too cool to contribute significant flux to the V band. This
is due to their large area. Illumination by the FUV continuum simply
cannot raise their temperature enough to make them contribute
significantly to the V band, let alone dominate its flux. The dashed
lines in Fig.\ref{fig5} show how little the illumination increases the
total BB disc flux at these very large radii.

The amount of FUV flux intercepted by large disc radii can be increased
by increasing the BB disc flaring, i.e. by increasing
$h_{out}/r_{out}$. The larger the outer disc scale-height, the more
illuminating flux the annulus intercepts, the greater the heating and
the more V band flux it contributes. We set $h_{out}/r_{out}=0.5$ but
this produces a negligible increase. The heating flux is simply spread
over too large an area. 

In order for a reprocessor located at $2\,ld$ ($R>1000\,R_g$) to contribute
significant V band flux it must have a small area. 
Clearly disc annuli are not suitable for this since the area of
the annulus is constrained to scale with its radius as $A(dr)\sim 2\pi
R dr$. An obvious source of small area reprocessors at large radii are
the broad line region (BLR) clouds. In the next section we investigate
the possibility that the observed optical lags are due to reprocessing
of the FUV emission, not by a BB disc, but by BLR clouds.

\subsection{Broad Line Region Clouds Reprocessing FUV Emission}

\begin{figure*} 
\centering
\begin{tabular}{l|r}
\leavevmode
\includegraphics[width=8cm]{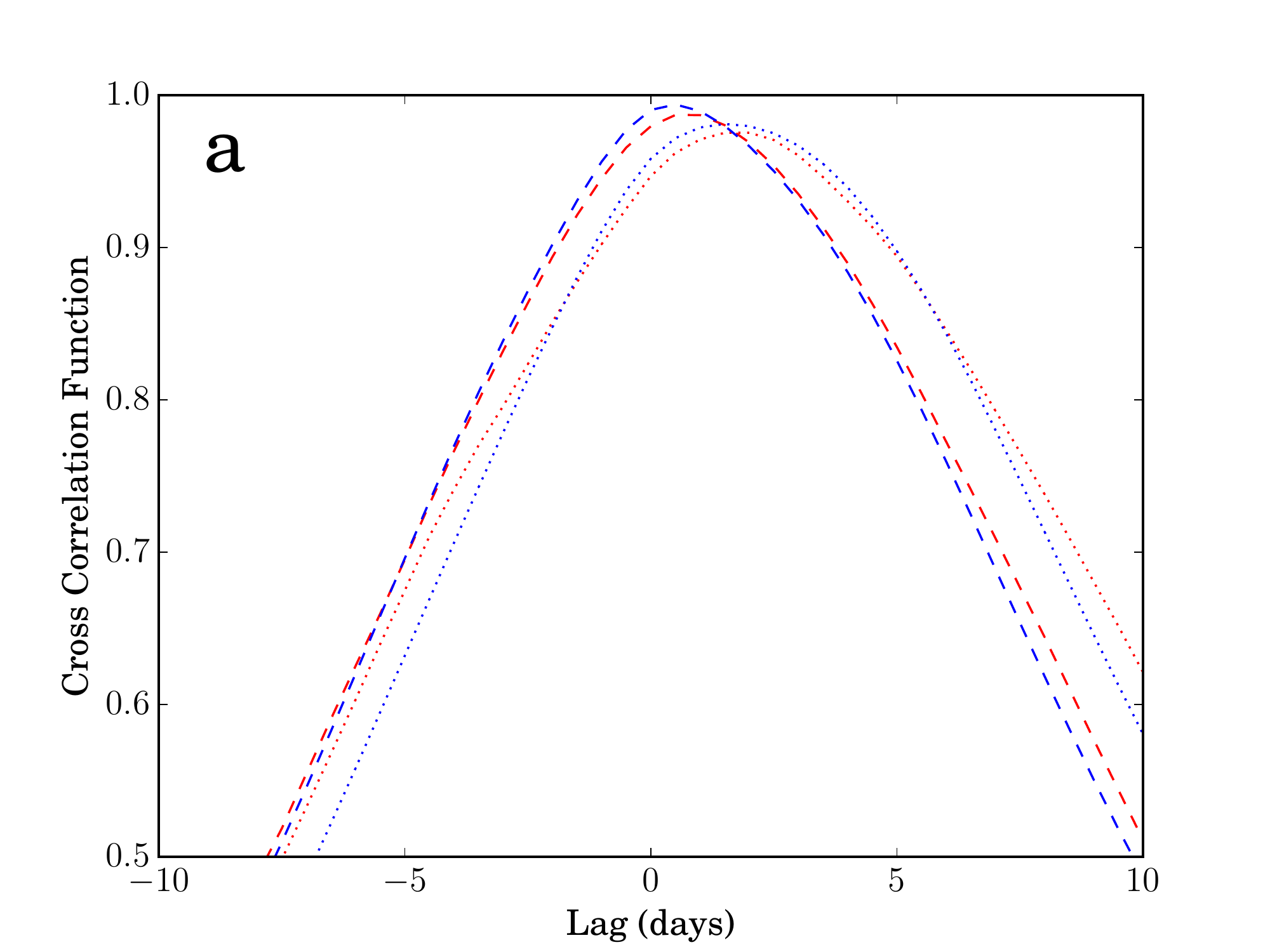} &
\includegraphics[width=8cm]{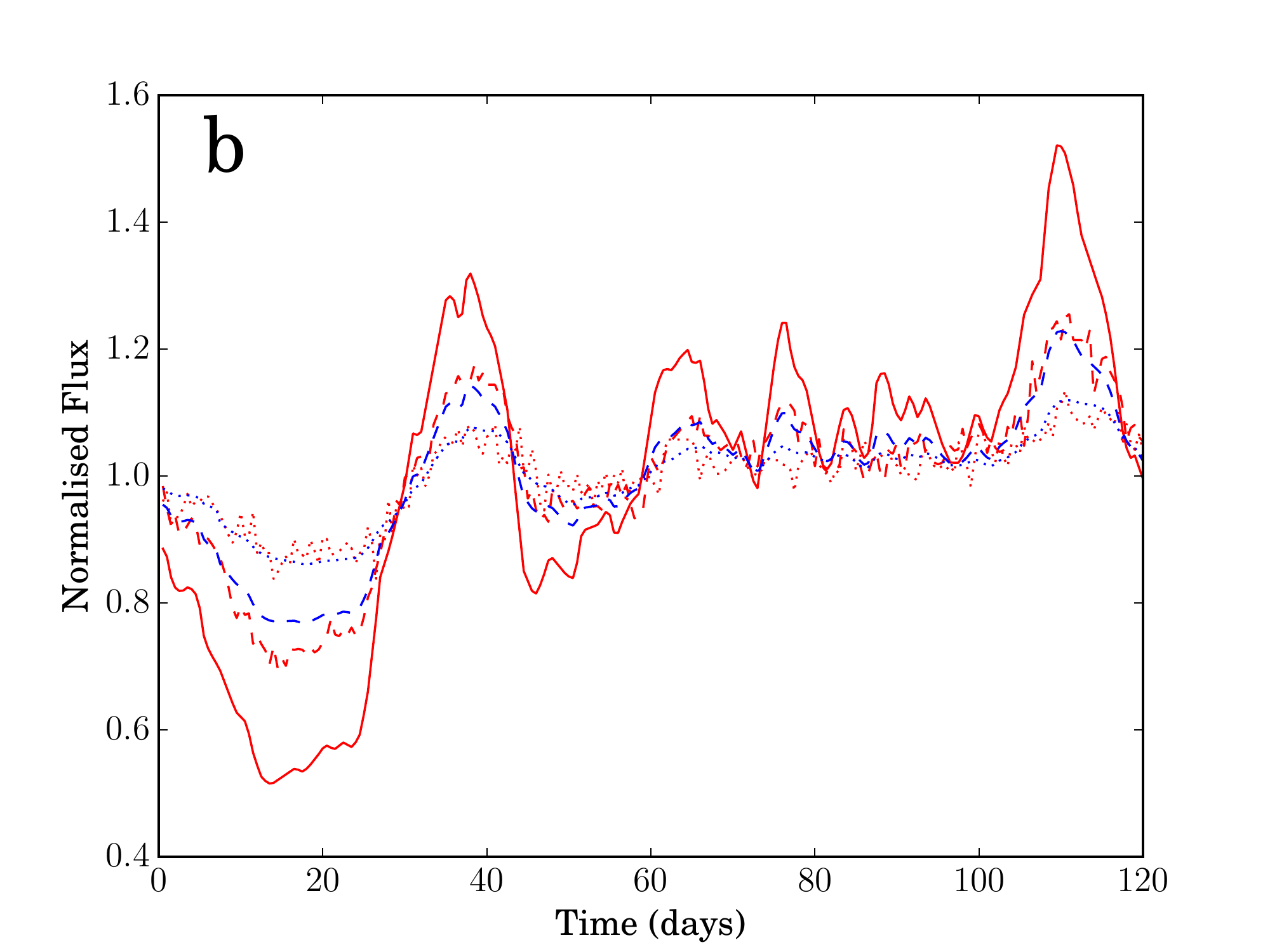} \\
\end{tabular}
\caption{Best fit thermal reprocessing model using observed CCFs and lightcurves to constrain reprocessor flux contributions and reprocessor properties. (a) Comparison of cross correlation functions. Dashed lines show CCF of UVW1 with respect to FUV, and dotted lines show CCF of V band with respect to FUV. Red lines show CCFs calculated using the observed lightcurves. Blue lines show the corresponding model CCFs using simulated UVW1 and V band lightcurves generated using the best fit component fractions listed in Table \ref{table2}. (b) Comparison of observed lightcurves (red) with simulated lightcurves (blue) generated using the component fractions listed in Table \ref{table2}. Solid line shows observed FUV lightcurve. Dashed lines show UVW1 lightcurves and dotted lines show V band lightcurves.}
\label{fig7}
\end{figure*}

BLR clouds absorb UV continuum emission and re-emit the energy as
optical lines/recombination continua. Clearly these do contribute to
the observed flux, and are lagged by the size-scale of the BLR. We
first explore if this contamination by the BLR can
influence the lags, as suggested by Korista \& Goad (2001). 

Mehdipour et al. (2015) show the UV/optical spectrum taken during the
campaign. We reproduce this in Fig.\ref{fig6}a (M. Mehdipour, private communication), 
with the continuum
bands superimposed. The UVW1 band contains a substantial amount of
blended Fe{\sc{II}} and Balmer continuum (Fig.\ref{fig6}a, magenta
line) as well as a broad Mg{\sc{II}} emission line (Fig.\ref{fig6}a,
blue line). The strongest line contribution to V band comes from 
the narrow [O{\sc{III}}] emission line (Fig.\ref{fig6}a, blue line), with a 
small contribution from the wing of H$\beta$. 

We conduct a simple test to determine whether this BLR line contamination
could explain the observed optical lags. The FUV is dominated by
continuum emission, so we assume the continuum component varies as the
FUV lightcurve, so that there are no real continuum lags. H$\beta$ 
lags the continuum in this source by roughly 15 days
(Peterson et al. 2002), though this does change with flux, spanning
4-20~days (Cackett \& Horne 2005; Bentz et al. 2010). 
We assume that all the BLR emission components (eg. Fe{\sc{II}}/Balmer blend, Mg{\sc{II}} and H$\beta$) are
a lagged and smoothed version of the FUV lightcurve, where the lag
and smoothing timescale is $15$\,d, while the narrow [O{\sc{III}}] emission line 
is constant on the timescale of our observations. Finally, we dilute this by the required
amount of constant host galaxy component (Mehdipour et al. 2015) to get
the full spectrum as a function of time. Integrating this over the
UVW1 and V bands gives the simulated lightcurves for this model. 
Fig.\ref{fig6}b shows the CCFs of these with respect to the FUV band lightcurve
(blue), in comparison with the CCFs of the real lightcurves (red),
with the UVW1 band as the dashed lines, and the V band as dotted. 
The red solid line shows the CCF of the FUV lightcurve with itself, 
i.e. the FUV autocorrelation function. 

Both model CCFs peak at zero, whereas the observed CCFs have peaks
offset from zero.  This is because the flux contribution from broad
lines is simply not large enough to shift the CCF peaks away from zero
in either UVW1 or V band. Furthermore, UVW1 contains more broad line
contamination than V band, and as a result the UVW1 model CCF (blue
dashed line) is more positively skewed than the V band CCF, which is
almost identical to the FUV ACF (compare blue dotted and red solid
lines). This is in clear contrast to the data, where the V band
lightcurve contains less line contamination than UVW1 and yet shows a
longer lag than UVW1. The UVW1 and V band lags therefore cannot be
explained through contamination by lagged broad line emission.

\section{Thermal Reprocessing}

So far, we have worked forwards from a geometric model of the spectrum
and its reprocessed emission, then calculated the resulting timing
properties and compared these to the observations. Now we take the
opposite approach. We begin by matching the timing properties of the
source (specifically the lightcurve amplitudes and CCFs) and use
these to infer the spectral components and then the geometry. 

We begin by matching the shape of the UVW1 and V band CCFs. 
The peak lag (i.e. the lag at which the CCF is a
maximum) of the observed UVW1 lightcurve with respect to
the observed FUV lightcurve is $\sim0.5-1$\,d. The CCF of the FUV
lightcurve with respect to itself (i.e. the FUV autocorrelation)
peaks at zero, as there is no lag between the two lightcurves. The
CCF of the FUV lightcurve lagged and smoothed by one day with respect
to the original FUV lightcurve will peak at $1$\,d. This is not the
only way a CCF peak at $1$\,d can be produced. If a lightcurve
consists of equal amplitude contributions from two lightcurves, one
of which is the original un-lagged FUV lightcurve and the other of
which is the FUV lightcurve lagged and smoothed by $2$\,d, then the
CCF of this composite lightcurve with respect to the original FUV
lightcurve will be the sum of the two CCFs --- the un-lagged FUV with
respect to itself and the $2$\,d-lagged FUV with respect to the
unlagged-FUV --- so that the resulting CCF will peak, not at $0$\,d or
$2$\,d, but at $1$\,d. More generally, any composite lightcurve CCF
will be the sum of the component lightcurve CCFs weighted by the
fraction of the total band flux coming from each lightcurve that is
correlated with the reference lightcurve. In this way, we can
constrain the flux contributions of variable components with different lags to
a given band by matching the peak and shape of the CCF.

We use sixteen variable component lightcurves: the
original FUV lightcurve, the FUV lightcurve lagged and smoothed by
$1$\,d, the FUV lightcurve lagged and smoothed by $2$\,d, the FUV lightcurve lagged and smoothed by $3$\,d, e.t.c, up to a maximum lag of $15$\,d. We then combine these component lightcurves to simulate model UVW1 and V band lightcurves and then compare the CCFs of these model lightcurves to the observed CCFs. We systematically adjust the component lightcurve contributions and select the fractional contributions which produce the smallest difference ($\Delta$) between model and observed CCFs, where $\Delta=\sum_{\tau =-20d}^{\tau =+20d}\lvert CCF_{UVW1,obs}(\tau)-CCF_{UVW1,model}(\tau)\rvert+\sum_{\tau =-20d}^{\tau= +20d}\lvert CCF_{V,obs}(\tau)-CCF_{V,model}(\tau)\rvert$. Since UVW1 and V band are spectrally close (and the blackbodies we will fit to the components are broad in comparison), we require both bands to contain the same lagged components (although the fractional contributions of these components in each band will differ). 

We find the best fitting model under these constraints
requires UVW1 and V band to contain a contribution from the original
FUV lightcurve, plus a contribution from the FUV lightcurve lagged and
smoothed by $6$\,d. Table \ref{table2} lists the corresponding
fractional contributions ($f_{FUV}$ and $f_{FUV-6d}$), while
Fig.\ref{fig7}a shows the resulting model CCFs (blue), compared to the
observed CCFs (red), where UVW1 CCFs are shown with dashed lines and V
band CCFs with dotted lines. The model V band CCF peaks at $1.5-2$\,d,
in agreement with the data. The UVW1 CCF peaks at $0.5$\,d, again in
agreement with the data.

Matching the CCFs allows us to constrain the
lagged, i.e. variable, flux contributions to the UVW1 and V bands. If
the UVW1 and V bands contained only these variable components
then they would have equal amplitude fluctuations. The observed FUV,
UVW1 and V band lightcurves shown in Fig.\ref{fig7}b in solid, dashed
and dotted red lines, respectively, clearly do not have equal
amplitude. The amplitude decreases with increasing wavelength,
implying the fluctuations are being increasingly diluted by a constant
component. Matching the amplitude of our simulated UVW1 and V band
lightcurves to the observations allows us to constrain the fractional
contribution of this constant component ($f_c$) to each band.

In reality this constant component has two possible sources: intrinsic
BB disc flux and flux from the host galaxy. The spectral decomposition in
Mehdipour et al. (2015) (Fig.\ref{fig6}a) shows that the contribution
of the host galaxy to the UVW1 band is negligible, hence we can assume
all the constant flux in the UVW1 band is supplied by intrinsic BB disc
emission (i.e. $f_d=f_c$, $f_g=0$). In contrast, the spectrum in
Mehdipour et al. (2015) shows that the flux contributions of the 
intrinsic BB disc continuum and the host galaxy are roughly equal in the
V band, i.e. $f_d=f_g=f_c/2$. Each band therefore has contributions
from a maximum of six spectral components, with the proportions of
each listed in Table \ref{table2}.

Having established the flux contributions of each temporal component
to each band, we can now begin finding spectral components that can
provide these flux levels in each band.

\begin{table}
\begin{tabular}{lcc}
\hline
 & UVW1 & V band \\
 \hline
$f_{FUV}$ & 0.395 & 0.189 \\
$f_{FUV-6d}$ & 0.105 & 0.111 \\
$f_c$ & 0.500 & 0.700 \\
$f_d$ & 0.500 & 0.350 \\ 
$f_g$ & 0.000 & 0.350 \\
\hline
\end{tabular}
\caption{Lightcurve fractions from model fits to the observed UVW1 and V band CCFs shown in Fig.\ref{fig7}. $f_{FUV}$ is the fraction of total band flux contributed by un-lagged FUV lightcurve. $f_{FUV-6d}$ is the fraction of total band flux contributed by the FUV lightcurve lagged and smoothed by 6 days. $f_c$ is the fraction of total band flux that is constant. There are two possible sources of constant flux: intrinsic BB disc emission ($f_d$) and the host galaxy ($f_g$), such that $f_d+f_g=f_c$.}
\label{table2}
\end{table}

\begin{figure} 
\centering
\begin{tabular}{l|r}
\leavevmode
\includegraphics[width=8cm]{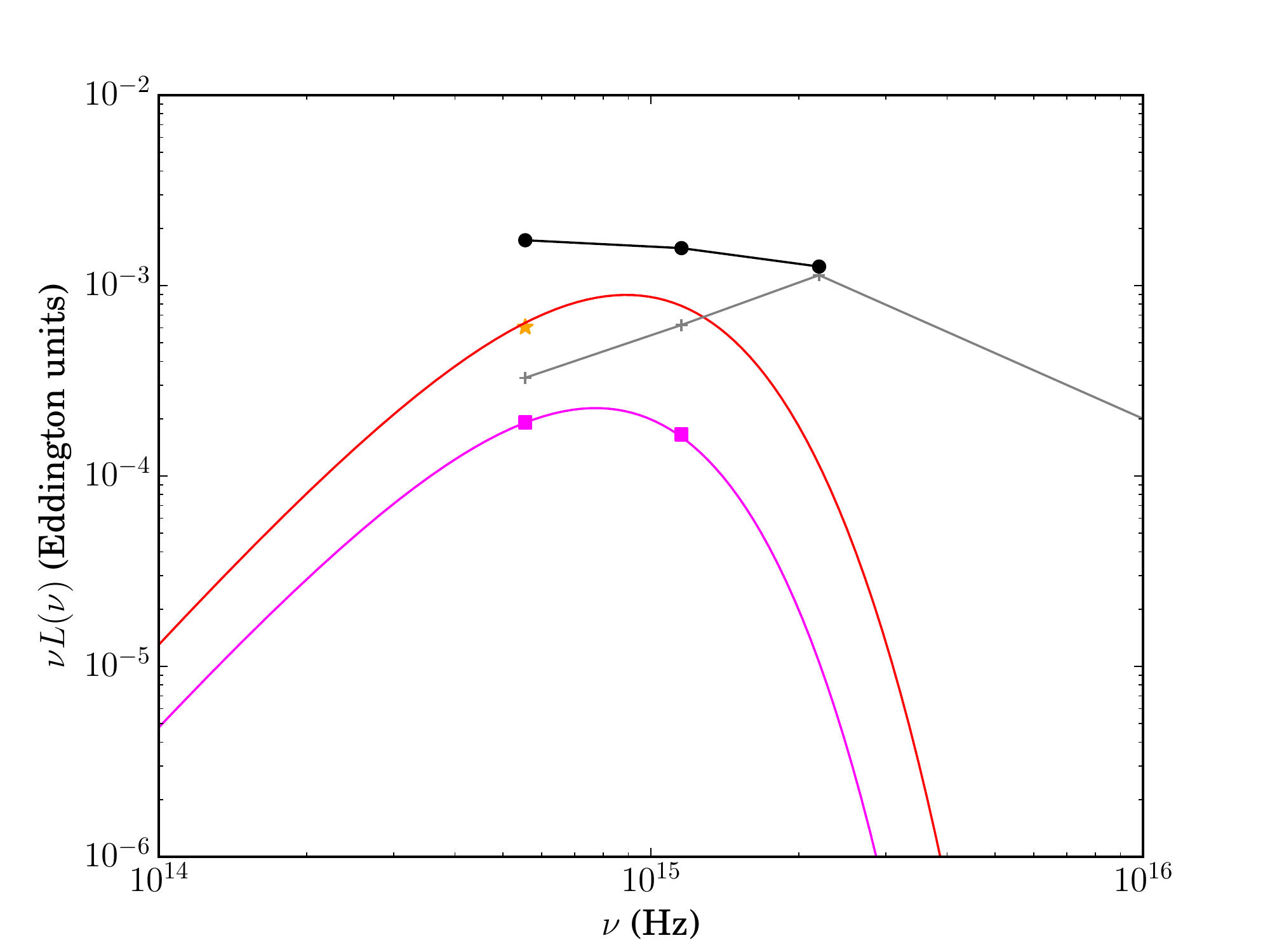} \\
\end{tabular}
\caption{Spectral decomposition inferred from the component fractions listed in Table \ref{table2}, which are shown to reproduce the observed timing properties of NGC 5548 in Fig.\ref{fig7}. Red line shows intrinsic (constant) emission from a BB disc truncated at $r_{in}=200$, with $r_{out}=300$. Black points, from left to right, show total flux in the V, UVW1 and FUV bands respectively. Orange star shows galaxy flux contribution to V band. Grey crosses mark the inferred flux levels of the spectral component varying as the FUV lightcurve. Magenta squares mark the inferred flux levels of the variable component with a $6\,$d lag with respect to the FUV lightcurve, while the magenta line shows the corresponding blackbody spectrum that matches these flux levels. The distance, temperature and covering factor of this blackbody component are: $R=3240\,R_g$, $T=9600$\,K and $f=0.002$.}
\label{fig8}
\end{figure}

We begin with intrinsic BB disc emission. We introduce a standard BB disc
(red line, Fig.\ref{fig8}), truncated at $r_{in}=200$ with $\log L/L_{Edd}=-1.4$, as is
required by the energetically constrained 
spectral decomposition shown in Fig.\ref{fig4}a. This
is the sole provider of constant flux in the UVW1 band. Since we know
the fractional contribution of constant flux to the UVW1 band
($f_c=f_d=0.5$; Table \ref{table2}), this tells us the total UVW1 band
flux is $F_{UVW1}=F_d(v_{UVW1})/f_d$ (central black point,
Fig.\ref{fig8}). From Fig.6a\&b in Mehdipour et al. (2015) we can
derive the ratio of dereddened FUV flux -- UVW1 flux ($vF(v)_{FUV}
\sim 0.8vF(v)_{UVW1}$), and the ratio of (deredenned but still
including host galaxy) V band flux --- UVW1 flux ($vF(v)_{Vband} \sim
1.1vF(v)_{UVW1}$). From our total UVW1 band flux we can therefore
calculate the total FUV and total V band flux (right hand and left
hand black points, Fig. \ref{fig8}). We know from Table \ref{table2}
that the fraction of intrinsic BB disc flux in the V band is $f_d=0.35$,
i.e. $F_d(v_{V band})=f_dF_{V band}$, which requires
$r_{out}=300$, in order to not over-predict the V band disc flux.

Using the total band fluxes and the fractional contributions in Table
\ref{table2} we can similarly constrain the UVW1 and V band flux
levels of the $6$\,d lagged component (magenta squares,
Fig.\ref{fig8}). We then construct a blackbody spectrum that can supply
these flux levels, shown in Fig.\ref{fig8} by the magenta line. 
A blackbody spectral component has two parameters: its
temperature and its area. The temperature defines the frequency of the
blackbody peak, while the temperature and emitting area together determine the luminosity of the blackbody. We adjust the temperature until the peak is placed
such that we can match the ratio of that component's UVW1 to V band
flux. We then adjust the area of the blackbody reprocessor to give
the blackbody the appropriate luminosity. 

The grey crosses in Fig.\ref{fig8} show the flux level of the
component which varies as the observed FUV lightcurve (the flux level
in the FUV band is found by subtracting the flux contributions of the
other components from the total FUV flux). The spectral shape of this
component is qualitatively similar to the Comptonised disc component shown
in Fig.\ref{fig4}a. Although the un-lagged component dominates in the FUV
band, there is some contamination from the lagged blackbodies. We
recalculate the FUV lightcurve including this contamination to check
that it does not affect the CCF peak lags and find its effect is
negligible.

Separating the variability of the source into its different components
and constraining the contribution of these components to different
bands, has allowed us to constrain the spectrum of the
reprocessor. Assuming this reprocessor is a blackbody emitter,
allowed us to constrain its temperature ($T=9600$\,K) and emitting area ($A=3.04\times 10^{30}$~cm$^2$
implying a luminosity $L_{BB}=1.37\times 10^{42}$ ergs~s$^{-1}$). The radial
location of the reprocessor is constrained by its lag time
i.e. it must be located six light days from the central source, which is $3240\,R_g$ for
NGC 5548 with $M=3.2\times10^7\,M_\odot$. Hence we can derive the
covering factor, $f_{cov}$, of the reprocessor, since $f_{cov}=A/(2\pi R^2)=0.002$. This is tiny and further underlines why the reprocessor cannot be a disc, which has a huge area at these radii. 

A possible source of small area reprocessors at large radii could be
dense clumps originating in a dust-driven disc wind just like the
broad line emitting clouds (Czerny et al. 2015), which are too dense
to emit broad lines so instead reprocess the illuminating flux as
thermal emission (see also Lawrence 2012). Knowing the covering fraction, it is clear that this 
material only intercepts 0.2\% of the illuminating flux. 
This is far too small to produce the observed luminosity required for the lagged component, given the observed bolometric luminosity of the source ($\sim2\times10^{44}$; Mehdipour et al. 2015). 

Hence we can rule out the lag originating from the reprocessing of 
irradiating flux, where the reprocessing makes blackbody radiation
and the lag is from the light travel time. Such a model fails on energetic grounds. 
A blackbody reprocessor at large radii
($\sim2000-3000\,R_g$, as required by the lag times) must have a small
area (ie. covering fraction), if it is to emit at a high enough
temperature to contribute significant flux to UVW1 and V band, which means it 
cannot intercept
sufficient illuminating flux to reprocess and heat it to this required
temperature. 

It seems more likely that the lag is not simply from light travel time delays, but is instead
lengthened by some response of the disc structure to the changing illuminating flux.

\subsection{An Alternative Explanation for the UV/Optical Lags}

So far we have assumed the FUV regions of the puffed-up,
optically thick, Comptonised disc region illuminate some separate
reprocessor --- either a standard BB disc or optically thick
BLR clouds. Having shown that neither reprocessor can produce the
observed lags, we suggest that perhaps light travel time lags from an
illuminating source to an external reprocessor are not involved at
all; perhaps the lags instead represent the lag time for the BB disc vertical structure 
to respond to changes in the FUV illumination. 

In this scenario, the hard X-rays heat the inner (soft
X-ray emitting) edge of the puffed-up Comptonised disc and this causes a
heating wave, which dissipates outwards. From examination of the observed
lightcurves (Fig.\ref{fig3}) we know this heating wave must quickly
lose the high frequency power of the hard X-rays, and it must also
include some intrinsic fluctuations produced within the Comptonised disc itself. We speculate that an increase in the hard X-ray flux produces a stronger heating wave, which dissipates outwards through the Comptonised disc. When this heating wave 
reaches the outer edge of the Comptonised disc, this increases the FUV illumination 
of the surrounding BB disc. This illumination is concentrated on the innermost BB disc radii 
adjacent to the Comptonised disc. These BB disc radii respond to the increase in illumination by 
expanding upwards, becoming less dense and less able to thermalise, so they may switch from emitting 
BB radiation to emitting via optically thick Compton --- the Comptonised disc region has essentially expanded outwards. When the hard X-ray flux decreases, there is less heating of the Comptonised disc region and perhaps 
its outer radii can then cool and return to emitting BB. The Comptonised disc region is effectively breathing in and out in response to the X-ray heating of its inner edge. 
We suggest this expanding and contracting of the puffed-up Comptonised disc region, ie. this movement of the transition radius between Comptonised disc and BB disc, is then the cause of the interband UV--optical lags. The lag times should therefore reflect the response time of the disc vertical structure to changing irradiation. 

The fastest response timescale of the disc is the dynamical timescale.
This sets both the orbital timescale, and the timescale on which the vertical structure 
responds to loss of hydrostatic equilibrium. The
latter seems more appropriate for the physical mechanism we envisage, 
but this is 32 days for a mass of $3.2\times 10^7M_\odot$
at $200\,R_g$. It is only as short as 6 days for $70R_g$. 
Hameury et al. (2009) calculate the effect of the disc instability for an AGN, but their illumination geometry is for a central source rather than the larger scale height FUV illumination envisaged here. 
More detailed simulations
are required to see if such a mechanism is feasible and if so, how to 
determine the radius more precisely from the timescale.

\section{Conclusions}

We have built a full disc reprocessing model in an attempt to
simulate the simultaneous multi-wavelength lightcurve data presented
by Edelson et al. (2015), assuming the UV/optical variability seen in
NGC 5548 is due to reprocessing of higher energy radiation by a BB
accretion disc. This higher energy radiation is traditionally assumed
to be the hard X-ray power law, produced at small accretion flow
radii. We find that reprocessing of the hard X-rays by a standard BB
accretion disc cannot replicate the observed UV/optical lightcurves
or their lags with respect to the illuminating X-rays. Specifically
the simulated lightcurves reproduce too much of the hard X-ray high
frequency power and the light travel lag times are too short.

One obvious answer to increase the amount of smoothing and increase
the light travel lag times would appear to be to make the BB accretion
disc larger. The first reason this is not an option is a constraint
from the spectral energy distribution of NGC 5548. We have limited the
size of our model accretion disc to the self-gravity radius. We could
relax this condition and allow the BB disc to extend to larger radii, but
this would cause the disc spectrum to extend to lower energies. By
contrast, the observed spectrum of NGC 5548, as shown by Mehdipour et
al. (2015), clearly does not show such emission. The peak UV/optical
emission occurs in the UVW1 bandpass and the emission below this
energy rapidly drops off. With an outer disc radius at the
self-gravity radius, we already slightly overpredict the lowest energy
fluxes. However, we modelled this directly, and find that the radii
required to smooth the reprocessed hard X-ray lightcurve to a level
matching the data is so large that the resulting lag times are too
long to be compatible with the observations.

We are then driven to a scenario where the reprocessor cannot directly
see the hard X-rays. This requires some source of material with
sufficient scale-height that it can block the optical emitting
regions' view of the hard X-rays. The obvious candidate for this is
the extra component required to fit the soft X-ray excess.

There is much debate over the physical origin of this component. 
We envisage a scenario where it is produced in an optically
thick Comptonising region at larger radii than the central hard
X-rays. UV/FUV emission produced in this region lifts material out of
the plane of the accretion flow, where it is illuminated by the hard
X-ray flux, which overionises the material. The resultant loss of UV
opacity means it falls back down, resulting in a region with
scale-height sufficient to prevent the hard X-rays illuminating the
outer BB disc. The soft X-ray emission could come
from the inner radii of this Comptonised disc region, while the lowest energy FUV
emission could come from its largest radii. 

One problem with complete shielding is that the observed FUV and hard
X-ray lightcurves {\em are} significantly correlated, though the
correlation is quite poor. One possibility to incorporate some
feedback between the two regions is if hard X-rays illuminate the
inner edge of the Comptonised disc region and these fluctuations then
dissipate outwards through the Comptonised disc until they reach the outer
FUV emitting regions, which then illuminate the surrounding BB disc. It is then
dissipation of the hard X-ray fluctuations through the Comptonised disc 
component that causes the loss of high frequency power, not light
travel time smoothing. This dissipation process has to be
extremely fast, however the viscous timescale in the Comptonised 
section of the disc should be faster than in a standard Shakura-Sunyaev
BB disc. 

We show that a model where the FUV emission (represented by the Hubble
band lightcurve) provides the illumination gives a much better match
to the shape of the observed optical lightcurves. 
However, our model
response is still too fast at the longest wavelengths. The observed V
band lag in NGC 5548 is $\sim2$\,d behind the FUV emission, which
requires the reprocessed V band flux to be emitted at radii
$R>1000\,R_g$ for this source. However BB disc annuli at these large
radii are too cool to contribute significant flux to the V band, which
is instead dominated by hotter emission from smaller BB disc radii with
shorter lag times. The heating effect of the illuminating flux makes
very little difference, as large disc annuli have enormous area. The
illuminating flux is simply spread over too large an area, so barely
changes the temperature of the annulus and certainly cannot heat the
annulus enough for it to contribute significant V band flux. In order
for the illuminating flux to heat a reprocessor at these large radii
enough to contribute to the V band requires the reprocessor to have a
small area.

We use the UV/optical lightcurves and cross correlation
functions to constrain the amount of reprocessed flux with different
lag times in the UVW1 and V bands and find a combination of unlagged
FUV lightcurve and FUV lightcurve lagged by 6d can fit both
bands. Assuming this 6d-lag reprocessed flux is blackbody emission,
we then estimated the temperature and covering factor of the blackbody reprocessor and find $T=9600$\,K and $f_{cov}=0.002$. This tiny
covering factor rules out the BB disc as the source of the reprocessed
emission. We consider the possibility that this emission may arise
from optically thick BLR clouds that are too dense to emit via line
emission so instead reprocess the FUV flux as thermal blackbody
emission. However this scenario is ruled out on energetic grounds ---
the inferred covering factor is too small for the reprocessor to
intercept sufficient illuminating flux to heat it to the required
temperature.

We conclude that the UV/optical lightcurves of NGC 5548 are not
consistent with reprocessing of the hard X-rays by a BB accretion disc,
but can instead be explained by reprocessing of the FUV emission
where the lag is not a light travel time. 
We propose the continuum lags of NGC 5548
are entirely generated by the `puffed-up' Comptonised disc region of
the accretion flow: the inner (soft X-ray emitting) edge of this
region is heated by the hard X-rays, producing heating waves which
dissipate outwards. The outer (UV/optical emitting) edge of the
Comptonised disc then expands and contracts, in both radius and height, in
response to the passage of these heating waves and it is this
behaviour which produces the continuum lags.  The model
still requires the presence of a standard outer Shakura-Sunyaev
BB accretion disc, but this is a mostly constant component.

Ultimately, whatever the origin of the lags in NGC 5548, the datasets
now available contain much more information than is encapsulated in a
single lag time measurement. We urge full spectral--timing modelling
of these data in order to extract all the new physical information on
the structure and geometry of the accretion flow which is now within
reach.

\section{Acknowledgements}

EG and CD acknowledge funding from the UK STFC grants
ST/I001573/1 and ST/L00075X/1. CD acknowledges Missagh Mehdipour for
multiple very helpful conversations and data, and Hagai Netzer for discussions
about thermalisation. We thank the referee for their detailed report which made us
assess the energetics of our original reverse engineered model with small clouds, and 
hence change the conclusions of our paper.

\label{lastpage}

\end{document}